\documentclass[11pt]{article}

\usepackage[hmargin=1.0in,vmargin=1.0in]{geometry}

\usepackage{setspace}
\usepackage{amsmath,amssymb,graphicx,color,amsthm,bm}
\usepackage[linesnumbered,ruled]{algorithm2e}
\usepackage{tikz}
\usepackage{dsfont}
\usepackage{mathtools}
\usetikzlibrary{shapes.geometric, arrows}
\usepackage{algorithmic}
\usepackage[english]{babel}
\usepackage[utf8]{inputenc}
\usepackage{mathrsfs}  
\usepackage{afterpage}
\usepackage{mathtools}
\usepackage{amsfonts}
\usepackage[colorinlistoftodos]{todonotes}
\usepackage{enumerate}
\usepackage[round]{natbib}
\usepackage{hyperref}
\usetikzlibrary{automata,positioning}
\usetikzlibrary{decorations.pathreplacing}

\newcommand{\excise}[1]{}

\usepackage{newfloat}
\DeclareFloatingEnvironment[name={Supplementary Figure}]{suppfigure}

\newcommand{\given}{\, |\,}

\newcommand\RR{\mathbb{R}}

\newcommand\ZZ{\mathbb{Z}}
\newcommand\EE{\mathbb{E}}

\newcommand{\cov}{\mathrm{Cov}}

\newcommand\blfootnote[1]{%
  \begingroup
  \renewcommand\thefootnote{}\footnote{#1}%
  \addtocounter{footnote}{-1}%
  \endgroup
}

\newtheorem{theorem}{Theorem}
\newtheorem{definition}{Definition}
\newtheorem{lemma}{Lemma}
\newtheorem{example}{Example}
\newtheorem*{example*}{Example}
\newtheorem{proposition}{Proposition}
\newtheorem{corollary}{Corollary}

\DeclareMathOperator\diag{diag}

\DeclareMathOperator\Id{I}

\DeclareMathOperator\rank{rank}
\DeclarePairedDelimiterX{\infdivx}[2]{(}{)}{%
	#1\;\delimsize\|\;#2%
}

\newcommand{\RNum}[1]{\uppercase\expandafter{\romannumeral #1\relax}}

\xdefinecolor{dukeblue}{rgb}{0.004,0.129,0.412}

\newcommand{\blind}{0}

\begin{document}
\def\spacingset#1{\renewcommand{\baselinestretch}%
		{#1}\small\normalsize} \spacingset{1}
\if0\blind
	{	
\title{Multi-group Gaussian Processes}
\author{\\[1ex]Didong Li$^{1,2}$, Andrew Jones$^{1}$, Sudipto Banerjee$^2$ and Barbara E. Engelhardt$^{1,3}$\\
{\em Department of Computer Science, Princeton University$^{1}$}\\
{\em Department of Biostatistics, University of California, Los Angeles$^{2}$}\\
{\em Gladstone Institutes$^{3}$}
}
\date{\vspace{-5ex}}
\maketitle
	} \fi
	
	\if1\blind
	{
		\bigskip
		\bigskip
		\bigskip
		\begin{center}
			{\Large \bf Multi-group Gaussian Process}
		\end{center}
		\medskip
	} \fi
	
	\bigskip

\begin{abstract}
    \noindent Gaussian processes (GPs) are pervasive in functional data analysis, machine learning, and spatial statistics for modeling complex dependencies. Modern scientific data sets are typically heterogeneous and often contain multiple known discrete subgroups of samples. For example, in genomics applications samples may be grouped according to tissue type or drug exposure. In the modeling process it is desirable to leverage the similarity among groups while accounting for differences between them. While a substantial literature exists for GPs over Euclidean domains $\RR^p$, GPs on domains suitable for multi-group data remain less explored. Here, we develop a multi-group Gaussian process (MGGP), which we define on $\RR^p\times \mathscr{C}$, where $\mathscr{C}$ is a finite set representing the group label. We provide general methods to construct valid (positive definite) covariance functions on this domain, 
    and we describe algorithms for inference, estimation, and prediction. We perform simulation experiments and apply MGGP to gene expression data to illustrate the behavior and advantages of the MGGP in the joint modeling of continuous and categorical variables.\blfootnote{Code for the model and experiments is available at \url{https://github.com/andrewcharlesjones/multi-group-GP}.}
    
\end{abstract}

\spacingset{1.45}

\section{Introduction}

Gaussian processes (GPs) are widely used in functional data analysis, machine learning, and spatial statistics due to their flexibility and expressiveness in modeling complex dependent data~\citep{rasm08, stein2012interpolation, gelfand2010handbook, cressie2011statistics, banerjee2014hierarchical}. For example, in nonparametric regression models, GPs are commonly used to model unknown arbitrary functions; in Bayesian contexts, they act as priors over functions \citep{ghosal2017fundamentals}. In spatial statistics, GPs are widely used to model spatial dependencies in geostatistical models and to perform spatial prediction or interpolation (``kriging'') \citep{matheron1963principles}. The GP framework may also be adapted to deal with discrete outcomes and applied to classification problems \citep{bernardo1998regression}. GPs are also being increasingly used in deep learning and reinforcement learning applications \citep{damianou2013deep, deisenroth2013gaussian}. 

A GP is determined by its covariance function, also known as the covariogram or Mercer kernel, which for statistical modeling purposes is specified to be a real-valued positive definite (PD) function $K: {\cal X} \times {\cal X} \to \RR$. Most commonly, GPs are defined over the Euclidean domain ${\cal X} = \RR^p$, where $p$ is the dimension of the input samples. However, in many cases, data sets admit a ``multi-group'' structure, where each sample belongs to one of $k$ known groups. For example, biological measurements might come from multiple tissue types or cell types~\citep{gtex2020gtex,regev2017science}; geospatial data might be collected from multiple locations defined by discrete demarcations, such as state or country borders~\citep{pan2020identifying}; and census data comes from people of different races, ethnicities, and genders~\citep{census2020}. Models built on Euclidean domains may be unsuitable for such data because they are not designed to capture the discrete nature of the groups. Although inference for GPs on non-Euclidean manifolds and graphs has recently attracted attention in statistics \citep{niu2019intrinsic,dunson2020diffusion,li2021fixed} and machine learning~\citep{NEURIPS2020_92bf5e62,borovitskiy2021matern}, the multi-group setting remains largely unaddressed. 

As a motivating example (which we will explore in depth in this paper), consider the data collected by the Genotype-Tissue Expression (GTEx) consortium~\citep{gtex2020gtex}. The GTEx data contain measurements from thousands of biological samples across hundreds of human donors, and these samples are collected from up to 52 tissue types for each donor. The measurements include gene expression and genotype profiles, along with a variety of additional metadata, including demographic variables and tissue health measurements. In this dataset, each tissue type constitutes a ``group.'' While each tissue type has a unique genomic environment, some tissue types are expected to be similar to one another. For example, a subset of the tissue types correspond to subregions of the brain, and another subset of tissue types correspond to subregions of the heart. In these cases, it may be beneficial to share information across similar tissue types when analyzing dependencies in the data.

For flexibly modeling $k$-group data, we seek a stochastic process over $\RR^p\times\mathscr{C}$ to drive the inference, where $\mathscr{C}=\{c_1,\cdots,c_k\}$ is a finite set representing the group labels. To begin, we identify three existing approaches for applying GPs to multi-group data: the Separated GP (SGP), the Union GP (UGP), and the Hierarchical GP (HGP)---all of which are defined over Euclidean domains. The SGPs and UGPs are straightforward applications of GPs that effectively ignore the group structure and either model the groups separately (SGP) or combine samples across groups and ignore the known structure (UGP). These ``separated'' and ``union'' modeling approaches are often used in practice for their simplicity and extend beyond the context of GPs~\citep{gtex2020gtex,tsherniak2017defining}. We briefly describe these three models below.
\begin{enumerate}
    \item The Separated GP (SGP) assumes independence across groups. Thus, the across-group correlation is set to zero: $K((x,c_i),(x',c_j))=0$ if $i\neq j$. The SGP is equivalent to modeling each group with a separate, independent GP.
    \item The Union GP (UGP) assumes the same dependencies within and across groups so the covariance function does not depend on the members of $\mathscr{C}$, i.e., $K((x,c_i),(x',c_j))=K_0(x,x')$. It is equivalent to modeling all groups jointly with a single GP.
    \item The Hierarchical GP (HGP) accommodates both across- and within-group dependencies, where all within-group and across-group dependencies are assumed to be identical. Here, $K((x,c_i),(x',c_j))=K_0(x,x')+\mathrm{1}_{\{c_i=c_j\}}K_1(x,x')$, where $K_0$ and $K_1$ are real-valued positive definite functions \citep{park2010hierarchical,hensman2013hierarchical}. 
\end{enumerate}
Each of the above models build covariance kernels based upon a standard GP over a Euclidean domain. While each of the above models offers a reasonable modeling solution for multi-group data, they offer limited flexibility. Specifically, the conditions implicit in the above models are restrictive and fail to model heterogeneity in the between-group dependencies. As a consequence, when the model is misspecified, the performance of these GPs will be unsatisfactory, especially when the overall sample size is small or the groups are unbalanced in terms of sample size.

In this manuscript we introduce a class of Multi-Group Gaussian Process (MGGP)  models as a flexible tool to model complex dependent grouped data. A specific contribution here is the construction of positive definite covariance functions defined over $\mathcal{X}\coloneqq\RR^p\times \mathscr{C}$ that will provide a valid specification for the MGGP over $\RR^p\times\mathscr{C}$. Specifically, our proposed MGGP framework will offer the following benefits compared to existing models: 
\begin{enumerate}
    \item The MGGP flexibly models multi-group heterogeneity, leveraging varying levels of similarity between groups;
    \item The MGGP is robust to model misspecification;
    \item The MGGP can make use of domain/prior/expert knowledge about which groups are expected to be similar to one another;
    \item The MGGP outperforms existing GPs, especially when the overall sample size is limited or when group-specific sample sizes are unbalanced;
    \item The MGGP does not depend on a sophisticated model; instead, the multi-group structure is encoded in the covariance function, which enables the seamless use of existing methods for estimation and inference in GPs;
    \item Given that the MGGP is a valid Gaussian process, it is compatible with the growing set of existing scalable Bayesian inferential methods for GPs \citep[see, e.g.,][and references therein for a review]{banerjee2017}.
\end{enumerate}

Much work in Bayesian statistics has focused on the challenge of joint modeling of continuous and categorical variables. One approach builds hierarchical models to partition the data based on each category and share strength across observations through the hierarchical structure~\citep{dunson2003bayesian,dunson2000bayesian,teimourian2015joint,ru2020bayesian}. Other approaches use additive effects for each category~\citep{schulam2015clustering}. More recent work uses Dirichlet process mixture models with both multinomial and Gaussian data likelihoods, which includes both additive mean effects and mixture components to effectively combine the above two strategies to model dependencies between continuous and categorical features \citep{murray2016multiple}. Our MGGP model adds to this literature by avoiding additive, hierarchical, and mixture models entirely, and instead explicitly modeling the dependencies within and among the two types of variables through a non-Euclidean covariance function with a flexible Gaussian process prior. The flexibility of the Gaussian process allows straightforward conditioning on multiple categorical partitions, offering a closed form conditional posterior. This is especially useful when some groups have extremely small sample sizes, and the posterior distribution can exploit dependencies between groups.

The paper is organized as follows. Section~\ref{sec: mggp_model} presents a model-based inferential framework for MGGP regression models. We turn to the construction of the MGGP using valid covariance functions on $\RR^p\times\mathscr{C}$ in Section \ref{sec: mggp} considering separable and inseparable functions in Sections~\ref{sec:separable_mggp}~and~\ref{sec:inseparable_mggp}, respectively, and show that the three aforementioned GPs for multi-group data are special cases in Section~\ref{sec:special_cases}. In Section~\ref{sec:experiments} we demonstrate our proposed method through simulation experiments and an application to a large gene expression data set. Section~\ref{sec:discussion} concludes the article with a brief discussion. Mathematical details, including all proofs, are presented in the Appendix.

\section{Multi-Group Gaussian Process Regression Models}\label{sec: mggp_model}
We envision a Multi-Group Gaussian Process (MGGP) regression model, where our dependent variable $y(x; c_j)$ is generated from a latent stochastic process over $\RR^{p} \times \mathscr{C}$ for inputs $x\in \RR^p$ and group $j$ as
\begin{equation}\label{eq: mggp_regression}
y(x; c_j) = \mu(x; c_j) + Z(x; c_j) + \epsilon(x; c_j)\;,\quad \epsilon(x; c_j) \stackrel{ind}{\sim} N(0, \tau_j^2)\;,
\end{equation}
where $\mu(x; c_j)$ is a mean function, $Z(x; c_j)$ is a zero-centered latent process and $\epsilon(x; c_j)$ is a zero-centered white-noise process capturing measurement error or fine-scale variation with group-specific variances. The mean function can be further modeled, if deemed appropriate, as $\mu(x; c_j) = f_j(x)^{\top}\beta_j$, where $f_j(x)$ is a $q_j\times 1$ vector of design variables possibly, but not necessarily, depending on $x$, and the $\beta_j$ variables are $q_j\times 1$ vectors of group-specific regression coefficients. This specification accommodates predictors or other explanatory variables that need neither be continuous nor reside within $\RR^{p}$.

The model in Equation \eqref{eq: mggp_regression} can be regarded as a semiparametric regression model with a parametric specification offered by the mean function and a nonparametric specification offered by the latent process. Our focus in this paper, however, is not so much on modeling $\mu(x; c_j)$, which can be built from customary parametric linear model specifications, as it is on the latent stochastic process $Z(x; c_j): \chi \longrightarrow \RR$, where $\chi = \RR^{p}\times \mathscr{C}$. We will specify $Z(\cdot; \cdot)$ to be a GP with zero mean and a covariance kernel $K((x; c_j), (x'; c_{j'})): \chi\times \chi \longrightarrow \RR$ so that $K(((x; c_j), (x'; c_{j'})) = \mbox{cov}(Z(x; c_j), Z(x'; c_{j'}))$ is a positive-definite covariance function. In general, we consider settings where data arise over a finite, possibly unbalanced, set of points $\{(x_i; c_j)\}$ for $i=1,2,\ldots,n_j$ and $j=1,2,\ldots,k$. Thus, each group can have a different number, $n_j$, of inputs. Given the covariance kernel, the realizations of the process over the finite set of points is the $n\times 1$ vector $Z = (Z_1^{\top},\ldots,Z_k^{\top})^{\top}$, where $n=\sum_{j=1}^k n_j$ and $Z_j = (Z(x_1; c_j),\ldots, Z(x_{n_j}; c_j))^{\top}$ follows a multivariate Gaussian distribution with an $n\times 1$ zero vector as mean and an $n\times n$ covariance matrix $K$, whose $(j,j')$th block is given by the $n_j\times n_{j'}$ matrix $K_{j{j'}}$ whose $(i,i')$ element is given by $K((x_i; c_j), (x_{i'}; c_{j'}))$ for $i=1,2,\ldots, n_j$ and $i'=1,2,\ldots,n_{j'}$.        

Equation~\eqref{eq: mggp_regression} leads to likelihood-based inference and can be extended to a Bayesian framework. Assuming, for elucidation purposes only, that $\mu(x; c_j) = f_j(x)^{\top}\beta_j$, our Bayesian MGGP model specifies the joint distribution of the parameters and the data as
\begin{equation}\label{eq: mggp_bhm}
    p(\{\tau_j^2\}, \theta, \{\beta_j\}) \times N(Z\given 0, K_{\theta})\times \prod_{j=1}^k\prod_{i=1}^{n_j} N(y(x_i; c_j)\given f_j(x_i)^{\top}\beta_j + Z(x_i; c_j), \tau_j^2)\;,
\end{equation}
where we now index the process covariance matrix by $\theta$ to denote parametric specifications for the covariance kernel, and $p(\{\tau_j^2\}, \theta, \{\beta_j\})$ is the prior on the model parameters. Inference on these parameters and the latent process proceeds by drawing samples from the posterior distribution $p(\{\tau_j^2\}, \theta, \{\beta_j\}, Z \given \{y(x_i;c_j)\}, \{f_j(x_i)\})$, which is proportional to Equation \eqref{eq: mggp_bhm}. 

Sampling from the joint posterior distribution including the process realizations $Z$ will be challenging due to the dimension of $Z$. Exploiting the Gaussian likelihood, we work with the collapsed likelihood after integrating out $Z$ from Equation \eqref{eq: mggp_bhm}, which yields the posterior distribution     
\begin{equation}\label{eq: mggp_collapsed}
   p(\tau, \theta, \beta \given y, F) \propto p(\tau, \theta, \beta) \times N(y \given F\beta, K_{\theta} + D_{\tau})\;,
\end{equation}
where $y$ is the $n\times 1$ vector of observations, $y(x_i; c_j)$, constructed analogous to $Z$, $F$ is an $n\times q$ block-diagonal matrix, $q=\sum_{j=1}^k q_j$, with $n_j\times q_j$ blocks $F_j = (f_j(x_1),\ldots,f_j(x_{n_{j}}))^{\top}$, $\beta = (\beta_1^{\top},\ldots,\beta_k^{\top})^{\top}$ is the $q\times 1$ vector of stacked regression coefficents, $\tau = \{\tau_j^2\}$ is the collection of measurement error variances and $D_{\tau}$ is the $n\times n$ diagonal matrix with $\tau_j^2I_{n_j}$ as $n_j\times n_j$ diagonal blocks. Markov chain Monte Carlo (MCMC) algorithms operate much more efficiently in sampling from Equation  \eqref{eq: mggp_collapsed} because of the reduced parameter space than from Equation \eqref{eq: mggp_bhm}. Once posterior samples of $\{\tau, \theta, \beta\}$ are collected, we recover the posterior samples for the latent process from $p(Z\given y, F) = \EE[p(Z\given \{\tau,\theta,\beta\}, y, F)]$, where the expectation $\EE[\cdot]$ is taken with respect to the posterior distribution in Equation \eqref{eq: mggp_collapsed}; Monte Carlo sampling will draw one $Z \sim p(Z\given \{\tau,\theta,\beta\}, y, F)$ for each posterior draw of $\{\tau,\theta,\beta\}$. This is straightforward because $p(Z\given \{\tau,\theta,\beta\}, y, F)$ is of the form $N(Mm, M)$, where $M^{-1} = K_{\theta}^{-1} + D_{\tau}^{-1}$ and $m = y - F\beta$, and the draws need to be made using only the post-convergence samples of $\{\tau,\theta,\beta\}$.         

To estimate the latent process at an unobserved input $x_0\in \RR$ for a given group $c_j\in \mathscr{C}$, we evaluate the Bayesian posterior predictive distribution
\begin{equation}\label{eq: mggp_predictive_given_group}
   p(Z(x_0; c_j) \given \{y(x_i;c_j)\}, \{f_j(x_i)\}) \propto \int p(Z(x_0; c_j) \given Z, \theta) \times p(Z, \{\tau,\theta,\beta\}\given y, F) dZ d\{\tau,\theta,\beta\}\;, 
\end{equation}
where we have used the conditional independence $p(Z(x_0; c_j) \given Z, \{\tau, \theta, \beta\}, y, F) = p(Z(x_0; c_j) \given Z, \theta)$ derived from the hierarchical model in Equation \eqref{eq: mggp_bhm}. We can sample from Equation \eqref{eq: mggp_predictive_given_group} by drawing one $Z(x_0; c_j) \sim p(Z(x_0; c_j) \given Z, \theta)$ for each drawn posterior sample of $Z$ and $\theta$, where $p(Z(x_0; c_j)\given Z, \theta)$ is Gaussian with mean $K_{\theta}((x_0; c_j); \cdot)^{\top}K_{\theta}^{-1}Z$, where $K_{\theta}((x_0; c_j); \cdot)$ is the $n\times 1$ vector with elements $K_{\theta}((x_0; c_j), (x_i, c_{j'}))$ for $j'=1,\ldots,k$ and $i=1,2,\ldots,n_{j'}$, and variance $K_{\theta}((x_0; c_j), (x_0; c_j)) - K_{\theta}((x_0; c_j); \cdot)^{\top}K_{\theta}^{-1}K_{\theta}((x_0; c_j); \cdot)$. Finally, to predict the response at $(x_0; c_j)$, we can sample from the predictive distribution $p(Y(x_0; c_j)\given y, F)$ by drawing one $Y(x_0; c_j) \sim N(f_j(x_0)^{\top}\beta_j + Z(x_0; c_j), \tau^2_j)$ for each posterior sample of $\{\beta_j, \tau^2_j\}$ and $Z(x_0; c_j)$, found as above. 

Our theoretical contribution includes the construction of valid positive-definite functions to serve as $K_{\theta}((x; c_j), (x'; c_{j'}))$. This is crucial for the above inferential framework as it ensures that the matrix $K_{\theta}$ in Equation \eqref{eq: mggp_bhm} will be positive definite for any finite set of distinct elements, observed or unobserved, in $\RR\times \mathscr{C}$. A key advantage of driving the inference through a latent process is the convenience of predictive inference for the process and the response at new inputs. A specific contribution of the proposed fully process-based framework is that it allows us to carry out predictive inference even for new unobserved groups. For example, if $c_j \in \mathscr{C}$ for some $j$ is a new group with no observations, sampling from the posterior predictive distributions of $Z(x_0; c_j)$ and $Y(x_0; c_j)$ can be executed as described above, possibly with appropriate modeling adjustments on the prior for $\tau_j^2$. Hence, we turn to the construction of valid MGGP covariance kernels.

It is worth pointing out the computational bottleneck arising from the dimension of $K_{\theta}$ in GP models for large data sets. There is, by now, a substantial literature on various approaches that build scalable models for massive data sets by building low-rank or sparsity-inducing processes \citep[see, e.g.][for expository treatments of such processes]{wikle_2011, banerjee2017, heatoncontest2019} out of any valid covariance kernel. While scalable processes is not the focus of the current manuscript, it is important to point out that our approach of constructing MGGPs using valid covariance kernels renders the resulting processes as ``scalable-ready'' since low-rank or sparsity-inducing MGGPs can be easily derived using current methods.

\section{Multi-Group Gaussian Processes}\label{sec: mggp}
We focus upon building MGGPs using positive-definite covariance kernels. We start with simpler separable specifications and proceed to derive richer and more flexible alternatives. 
\subsection{Separable multi-group GP}
\label{sec:separable_mggp}
We start with a simple case where the covariance function over $\RR^p\times\mathscr{C}$ is separable.
\begin{definition}
$K$ is said to be separable if $K((x,c_i),(x',c_j))=K_{\RR^p}(x,x')K_\mathscr{C}(c_i,c_j)$, where $K_{\RR^p}$ and $K_\mathscr{C}$ are over $\RR^p$ and $\mathscr{C}$.
\end{definition}
First, observe that $K$ is positive definite if and only if both $K_{\RR^p}$ and $K_\mathscr{C}$ are positive definite. Since GPs over $\RR^p$ have been thoroughly studied, we focus on covariance functions over $K_\mathscr{C}$, or GPs over a categorical set. Also, $\mathscr{C}$ being finite, any function on $\mathscr{C}\times \mathscr{C}$ is completely determined by $C\in\RR^{k\times k}: C_{ij}=K_\mathscr{C}(c_i,c_j)$. 
\begin{proposition}\label{prop:PD_cat}
$K_\mathscr{C}$ is positive definite if and only if $C$ is a positive definite matrix.
\end{proposition}
The above result implies that it will suffice to find a positive definite function on $\RR^p$ and a positive definite matrix $C\in\RR^{k\times k}$  
to construct a separable positive definite function on $\mathcal{X}$. Homogeneous kernels arise as a special case.
\begin{definition}
A function $K_\mathscr{C} :\mathscr{C}\times \mathscr{C}\to \RR$ is said to be homogeneous if $K_\mathscr{C}(c_i,c_j)=K_0( 1_{\{c_i\neq c_j\}})$ for some function $K_0$ on $\{0,1\}$.
\end{definition}

A homogeneous GP is completely determined by two scalars: $a\coloneqq K_\mathscr{C}(c_i,c_i)$, $b\coloneqq K_\mathscr{C}(c_i,c_j)$ where $i\neq j$, which represent the within-group correlation and across-group correlation, respectively. Without loss of generality, we assume $a=1$; otherwise, we can rescale $K_\mathscr{C}$. In this case, all within-group and between-group correlations are the same. A homogeneous GP is appropriate if we only need to distinguish pairs of observations that are in the same group from those in different groups, while the specific group identities are irrelevant. Equivalently, $K$ is homogeneous if $K$ is isotropic with respect to the discrete metric $d(c_i,c_j)=1_{\{c_i\neq c_j\}}$. 

\begin{corollary}\label{cly:PDZ2}
Let $K_\mathscr{C}$ be homogeneous, then $K_\mathscr{C}$ is positive definite if and only if $-\frac{1}{k-1}\leq b\leq 1$, where $b=K_\mathscr{C}(c_i,c_j)$ with $i\neq j$.
\end{corollary}
The inequality $-\frac{1}{k-1}\leq b\leq 1$ implies the across-group correlation should not dominate the within-group correlation, which is intuitively reasonable. 

Separable models provide computational benefits because the resulting covariance matrix for $Z$ can be expressed as a Kronecker product of $K_{\RR}$ and $K_{\mathscr{C}}$. However, separable covariance functions tend to have ``ridges" or discontinuities \citep{stein2005space} that can lead to poorer inference. They also assume that the same covariance structure ($K_{\RR^p}$) is retained for all groups, which seems unduly restrictive in terms of accommodating associations of the latent process for different pairs of inputs in $\RR^p\times \mathscr{C}$. Hence, we turn to inseparable cases in the next section.

\subsection{Inseparable multi-group GP}\label{sec:inseparable_mggp}
\subsubsection{Isotropic MGGP}
In order to discuss ``isotropic'' GPs on $\RR^p\times \mathscr{C}$, we need to endow $\mathscr{C}$ with more structure. To minimize assumptions, we introduce a metric $d$ on $\mathscr{C}$ so that $(\mathscr{C},d)$ is a metric space. 
\begin{definition}
Given two metric spaces $(\mathcal{Y},d)$ and $(\mathcal{Y}',d')$, a GP on $\mathcal{Y}\times \mathcal{Y}'$ is said to be semi-isotropic if $K((x_1,x_1'),(x_2,x_2'))=K_0(d(x_1,x_2),d'(x_1',x_2'))$. 
\end{definition}
Intuitively, a semi-isotropic GP is isotropic in each component $\RR^d$ and $\mathscr{C}$ separately. Isotropic functions imply semi-isotropic, but the other direction does not hold in general. In practice, $d$ is usually obtained by domain knowledge, including prior, exterior, or expert knowledge. For example, if $\mathscr{C}=\{\text{North Carolina, New Jersey, California}\}$, then the distance can be the geographical distance between the center of these states. As another example, if $\mathscr{C}$ contains a set of human tissue types, prior biomedical knowledge might lead to candidates for $d$; two tissue types from different parts of the same organ, say the brain, might be expected to be more similar to each other than a brain tissue and liver tissue. When $\mathscr{C}$ is a weighted graph, the graph distance serves as a valid metric \citep{bouttier2003geodesic}. If no domain knowledge is available, we suggest using a default noninformative distance: $d_{ij}=1-\delta_{ij}$, i.e., all groups are equidistant. 

Recall that the restriction of a GP to a subset of the original domain is again a GP \citep{rasm08}. Motivated by this fact, if we can isometrically embed $\mathscr{C}$ to an Euclidean space $\RR^{p'}$ by some mapping $\iota:\mathscr{C}\to\RR^{p'}$ such that
\[
d_{ij}\coloneqq d(c_i,c_j)=\|\iota(c_i)-\iota(c_j)\|,
\]
then an isotropic GP on $\RR^{p'}$ induces a GP on the image $\iota(\mathscr{C})$. Our next result creates a large family of semi-isotropic covariance functions on $\mathcal{X}=\RR^p\times\mathscr{C}$.
\begin{theorem}\label{thm:st}
Let $\iota$ be an isometric embedding from $(\mathscr{C},d)$ to $\RR^{p'}$. Then, if $\varphi: \RR_+\to \RR$ is a completely monotone function and $\psi:\RR_+\to \RR_+$ is a positive function with a completely monotone derivative, then:
\begin{equation}\label{eqn:spatialtemporal}
K((x,c_i),(x',c_j))=\frac{\sigma^2}{\left(\psi(\|\iota(c_i)-\iota(c_j)\|^2)\right)^{p/2}}\varphi\left(\frac{\|x-x'\|^2}{\psi(\|\iota(c_i)-\iota(c_j)\|^2)}\right).
\end{equation}
is a valid covariogram, where $\sigma^2>0$ is the spatial variance. In particular, if $d(c_i,c_j)=1-\delta_{ij}$ is the discrete metric, and if $\varphi: \RR_+\to \RR$ is a completely monotone function, then
\begin{equation}
K((x,c_i),(x',c_j))=\frac{\sigma^2}{\alpha^{\frac{p}{2}(1-\delta_{ij})}}\varphi\left(\frac{\|x-x'\|^2}{\alpha^{1-\delta_{ij}}}\right)
\end{equation}
is a valid covariogram, where $\sigma^2\in (0,1]$ is the spatial variance and $\alpha>0$ controls the interaction between $\RR^p$ and $\mathscr{C}$.
\end{theorem}

There are two remaining questions to fully define the covariance function: How do we determine $p'$? and how do we find the embedding $\iota$? In fact, the explicit formula of $\iota$ is not necessary to define the covariance function since only $d$ is involved in Equation \eqref{eqn:spatialtemporal}. Therefore, as long as $d$ is known, the above covariance functions are well-defined. If one is interested in the embedding $\iota$ itself, the following lemma answers the first question:
\begin{lemma}[\cite{maehara2013euclidean}]
 Let $G\in\RR^{k\times k}$ be the Gram matrix of $d$, that is,
 \[
 G_{ij}=\frac{1}{2}\left(d(c_1,c_i)+d(c_1,c_j)-d(c_i,c_j)\right)\;.
 \]
Then there exists an isotropic embedding $\iota:\mathscr{C}\to \RR^{p'}$ if and only if $G$ is positive semi-definite with rank at most $p'$.
\end{lemma}
For simplicity, we choose $p'=\rank(G)$ to reduce the dimension. For the second question, such an $\iota$ can be obtained using existing methods in closed-form, including a power transform \citep{maehara2013euclidean, hopkins2015finite}. If $\mathscr{C}$ has a graph structure, then graph embedding algorithms apply here, see \cite{deza2009geometry,frankl2020embedding} for more details. 

The solution for $\iota$ admits a simple form when $d$ is the discrete metric, that is, $d(c_i,c_j)=1-\delta_{ij}$. For the discrete metric, the Gram matrix is
$B=\begin{bmatrix}
0 & 0 & 0 & \cdots & 0 \\
0 & 1 & \frac{1}{2} & \cdots & \frac{1}{2}\\
0 & \frac{1}{2} & 1 & \cdots & \frac{1}{2}\\
\vdots & \vdots &\vdots& \ddots &\vdots \\
0 & \frac{1}{2} & \frac{1}{2} & \cdots & 1
\end{bmatrix}=\begin{bmatrix}
    0 & \bm{0}_{1\times (k-1)}\\
    \bm{0}_{(k-1)\times 1} & \widetilde{B}
\end{bmatrix},$
where $\widetilde{B} = \frac{1}{2}\Id_{k-1}+\frac{1}{2}\bm{1}_{(k-1)\times (k-1)} $ is full rank. As a result, $\rank{B}=k-1=p'$, so we can isometrically embed $\mathscr{C}$ to $\RR^{k-1}$. A natural embedding $\iota$ is to map each $c_i$ to the vertex of the $k-1$ standard simplex $\Delta_{k-1}\subset\RR^{k-1}$. Then the embedding $\iota$ is given by
\[
\iota: \mathscr{C}\to \Delta_{k-1}, c_i\mapsto \frac{1}{\sqrt{2}}\bm{e}_i-\frac{1}{(k-1)\sqrt{2}}\left(1+\frac{1}{\sqrt{k}}\right)\bm{1}_{k-1},~1\leq i\leq k-1,~ c_k\mapsto -\frac{1}{\sqrt{2k}}\bm{1}_{k-1}.
\]

\subsubsection{Candidate covariance functions for the MGGP}
Some candidates for completely monotone functions $\phi$ and positive functions with completely monotone derivatives $\psi$ are in Table \ref{table:completelymonotone} \citep{gneiting2002nonseparable}:

\begin{table}[h!]
\centering
\begin{tabular}{cc}
\hline 
 $\phi(t)$&  $\psi(t)$  
 \tabularnewline
\hline 
$\exp(-ct^\gamma)$ & $(at^\alpha+1)^\beta$ 
\tabularnewline
\hline 
$(2^{\nu-1}\Gamma(\nu))^{-1}(ct^{1/2})^\nu K_\nu (ct^{1/2})$ & $\log(at^\alpha+b)/\log b$ 
\tabularnewline
\hline 
$(1+ct^\gamma)^{-\nu}$ & $(at^\alpha+\beta)/(\beta(at^\alpha+1))$ \tabularnewline
\hline 
$2^\nu(\exp(ct^{1/2})+\exp(-ct^{1/2}))^{-\nu}$ & 
\tabularnewline
\hline 
\end{tabular}\label{table:completelymonotone}
\caption{Candidate functions for completely monotone functions $\phi$ and positive functions with completely monotone derivatives $\psi$. Here,  $a,c,\nu>0$, $b>1$, $0<\alpha,\beta,\gamma \leq 1$.}
\end{table}

Selecting a $\phi$ and a $\psi$ function from Table \ref{table:completelymonotone}, we obtain the following semi-isotropic covariance functions on $\mathcal{X}$, for more covariance functions, see Appendix \ref{sec:covs}:

\begin{eqnarray}
K((x,c_i),(x',c_j))&=&\frac{\sigma^2}{(a^2d_{ij}^2+1)^{p/2}}\exp\left\{-\frac{b^2\|x-x'\|^2}{a^2d_{ij}^2+1}\right\},\label{eqn:RBF}\\
K((x,c_i),(x',c_j))&=&\frac{\sigma^2}{(ad_{ij}+1)^{p/2}}\exp\left\{-\frac{b^2\|x-x'\|^2}{ad_{ij}+1}\right\},\label{eqn:RBF'}\\
K((x,c_i),(x',c_j))&=&\begin{cases}
\frac{\sigma^22c^{p/2}}{(a^2d_{ij}^2+1)^\nu(a^2d_{ij}^2+c)^{p/2}\Gamma(\nu)}\left\{\frac{b}{2}\left(\frac{a^2d_{ij}^2+1}{a^2d_{ij}^2+c}\right)^{1/2}\|x-x'\|\right\}^\nu \\ \quad\times K_\nu\left(b\left(\frac{a^2d_{ij}^2+1}{a^2d_{ij}^2+c}\right)^{1/2}\|x-x'\|\right) & x\neq x' \\
\frac{\sigma^2c^{p/2}}{(a^2d_{ij}^2+1)^\nu(a^2d_{ij}^2+c)^{p/2}} & x=x'
\end{cases},\label{eqn:matern}\\
K((x,c_i),(x',c_j))&=&
\frac{\sigma^2c^{p/2}}{(a^2d_{ij}^2+1)^{1/2}(a^2d_{ij}^2+c)^{p/2}}\exp\left\{-b\left(\frac{a^2d_{ij}^2+1}{a^2d_{ij}^2+c}\right)^{1/2}\|x-x'\|\right\}.\label{eqn:matern1/2}
\end{eqnarray}

In the equations above, $\sigma^2>0$ is the spatial variance, $a\geq0$ is the group similarity scale, $b\geq0$ is the feature scale, $c\geq0$ is the dependency scale, and $\nu>0$ is a smoothness parameter. The covariance functions in Equation \eqref{eqn:RBF} and Equation \eqref{eqn:RBF'} are analogues of the squared exponential or radial basis function (RBF). The covariance function in Equation \eqref{eqn:matern} is the analogue of the Mat\'ern covariance function. In particular, the covariance function in Equation \eqref{eqn:matern1/2} is the special case of Equation \eqref{eqn:matern} when $\nu=1/2$, so we can call it the exponential covariance function. Equation  \eqref{eqn:matern} becomes separable when $c=1$. 

\subsubsection{Stationary MGGP with $k=2$}
We now aim to weaken the isotropic condition described above. Since $(\mathscr{C},d)$ does not admit a natural group structure, we start with the simple case where $k=2$, which appears frequently in practice, including in data sets where the two groups are male/female, adults/children, treatment/control, etc. Note that by rescaling, we can always embed a binary metric space within $\RR$: $\iota(c_1)=0, ~\iota(c_2)=1$. Then any spatial-temporal covariogram induces an semi-isotropic covariance function for $\RR^d\times \mathscr{C}$. 

For the non-isotropic case observe that $\mathscr{C}$ can be identified with $\ZZ_2$ when $k=2$, an Abelian group. In this situation, $K$ is said to be stationary if $K((x,d),(x',l))=K_0(x-x',d-l)$. We will use $K$ instead of $K_0$ for simplicity, where $K$ is characterized by two covariance functions: $K_w = K(\cdot,0)$, the within-group covariance function, and $K_c=K(\cdot,1)$, the cross-group covariance function. It is clear that any covariance function on $\RR^p\times \ZZ_2$ determines two covariance functions on $\RR^p$. On the other hand, not all pairs of covariance functions on $\RR^p$ define a valid (PD) covariance function on $\RR^p\times \ZZ_2$. In order to construct a valid covariance function on $\RR^p\times \ZZ_2$, we need to find a sufficient condition for $K$ being PD.

\begin{theorem}\label{thm:BochnerZ2}
Let $K_w$ and $K_c$ be two PD functions on $\RR^p$ with spectral densities $\rho_w$ and $\rho_c$ such that
\[
K(x,0)=K_w(x)=\int_{\RR^p}e^{-2\pi i\omega x}\rho_w(\omega)d\omega,~~K(x,1)=K_c(x)=\int_{\RR^p}e^{-2\pi i\omega x}\rho_c(\omega)d\omega\;.
\]
Then, $K(x,l)=\begin{cases} 
K_w(x) & l=0\\ 
K_c(x) & l=1 
\end{cases}$ is PD on $\RR^p\times \ZZ_2$ if and only if $\rho_w\geq \rho_c$.
\end{theorem}

\begin{example}
Recall the RBF covariance function in Equation \eqref{eqn:RBF}:
\[
K(x,l)=\frac{\sigma^2}{(a^2l^2+1)^{p/2}}\exp\left\{-\frac{b^2\|x\|^2}{a^2l^2+1}\right\}.
\]
The two spectral densities are given by
\begin{eqnarray*}
\rho_w(\omega) &=& \sigma^2\left(\frac{\pi}{b^2}\right)^{\frac{p}{2}}\exp\left\{-\frac{\pi^2 \|\omega\|^2}{b^2}\right\}, \\
\rho_c(\omega) &=& \sigma^2\left(\frac{\pi}{b^2}\right)^{\frac{p}{2}}\exp\left\{-\frac{\pi^2(a^2+1) \|\omega\|^2}{b^2}\right\},
\end{eqnarray*}
where $\rho_w\geq \rho_c$.
\end{example}

The stationary MGGP assumes homogeneity in the within-group correlation; however, for certain applications, heterogeneity exists. To account for this heterogeneity, we consider a weaker version, which we call the semi-stationary MGGP. A semi-stationary MGGP is stationary in $\RR^p$ but not in  $\mathscr{C}$. 
\begin{definition}
$K$ is said to be semi-stationary if $K((x,c_i),(x',c_j))=K_0(x-x',c_i,c_j)$ where $K_0$ is defined on $\RR^p\times \mathscr{C}\times \mathscr{C}$.
\end{definition}
This semi-stationary assumption is suitable for applications where groups are expected to have different within-group correlations, but the GP is stationary once the group is fixed. For semi-stationary MGGPs, $K$ is determined by $K_{0}(x)=K(x,0,0)$,  $K_c(x)=K(x,0,1)=K(x,1,0)$ and $K_{1} = K(x,1,1)$, where $K_0\neq K_1$ in general, otherwise $K$ becomes stationary. 

\begin{theorem}\label{thm:BochnerZ2Z2}
Let $K_0$, $K_c$ and $K_1$ be PD functions on $\RR^p$ with spectral densities $\rho_0$, $\rho_c$, and $\rho_1$. Then $K(x,l,l')=\begin{cases} 
K_0(x) & l=l'=0\\ 
K_c(x) & l+l'=1  \\
K_1(x) & l=l'=1
\end{cases}$ is PD on $\RR^p\times \ZZ_2$ if and only if $\rho_0\rho_1\geq \rho_c^2$.
\end{theorem}

Data sets with more than two groups are ubiquitous in scientific applications. For example, in a biomedical setting, data may be collected from multiple human tissue types; from people of different race and ethnicity; and from varying age groups. Hence, we generalize the above theory to $k>2$ groups. The main difficulty here is that $\mathscr{C}$ does not admit a natural group structure for $k>2$. A straightforward solution would be to identify $\mathscr{C}$ with $Z_{k}$, but the modular structure of $Z_k$, that is, $1-0=2-1=\cdots k-1-(k-1-1)\neq k-(k-1)$, is not satisfied in practice. Hence, Bochner's Theorem, which fully characterizes positive definite functions on locally compact Abelian groups, is not applicable. 

Motivated by the proof of \autoref{thm:BochnerZ2Z2}, we observe that this result is formally identical to an application of Cram\'er's Theorem to bivariate GPs.
\begin{lemma}[\cite{cramer1940theory}]\label{lem:cramer2}
Let $\widetilde{K}:\RR^p\to \RR^{2\times 2}$ be a function with spectral density $\widetilde{\rho}_{ij}$. Then, $\widetilde{K}$ is positive definite, hence defines a bivariate GP on $\RR^p$, if and only if $\widetilde{\rho}(\omega)=\{\widetilde{\rho}(\omega)\}_{i,j=1}^2$ is positive semi-definite for any $\omega\in\RR^p$.
\end{lemma}
That is, \autoref{thm:BochnerZ2Z2} draws an equivalence between Bochner's Theorem on $\RR^p\times \ZZ_2$ and Cram\'er's Theorem on $\RR^p$ with bivariate outputs. This insight is crucial, as it enables us to leverage the large literature on bivariate GPs to construct two-group GPs. Furthermore, given that Cram\'er's Theorem holds for a general $k$-variate GP, this observation suggests a general theory for an arbitrary number of groups with $k>2$, which we explore below.

\subsubsection{Stationary MGGP with arbitrary $k$}
We first draw similarities between the MGGP with $k$ groups and $k$-variate GPs. Recall that a Gaussian $k$-variate random field $\widetilde{Z}$ on $\mathcal{Y}$ is characterized by its cross-covariance function $\widetilde{K}:\mathcal{Y}\times \mathcal{Y}\to\RR^{k\times k}$:
\[
\cov(\widetilde{Z}(x),\widetilde{Z}(x'))=\widetilde{K}(x,x').
\]
\begin{theorem}\label{thm:MGGPkGP}
Let $\mathscr{G}$ be the space of all Gaussian random fields on $\mathcal{Y}\times\mathscr{C}$, where $\mathscr{C}=\{c_1,\cdots,c_k\}$ and $\mathscr{V}$ is the space of all Gaussian $k$-variate random fields on $\mathcal{Y}$. Then 
\[
\Phi:\mathscr{G}\to\mathscr{V},~(\Phi(Z))_i(x) \coloneqq Z(x,c_i), ~\forall Z\in \mathscr{G}
\]
is a bijection, and its inverse $\Phi^{-1}$ is given by 
\[
\Phi^{-1}:\mathscr{V}\to\mathscr{G},~(\Phi^{-1}(\widetilde Z))(x,c_i) = \widetilde{Z}_i(x),~\forall \widetilde{Z}\in\mathscr{V}.
\]
The correspondence between the covariance function of $Z$ and the cross-covariance function of $\widetilde{Z}$ is given by 
\[
K((x,c_i),(x',c_j))=\widetilde{K}(x,x')_{ij}.
\]
\end{theorem}
This shows that in order to construct a $k$-group MGGP it will suffice to construct a $k$-variate GP, and vice versa. For our purposes, existing constructions of multi-variate GPs can be applied \citep[see, e.g.,][ for possible ideas.]{gneiting2010matern, apanasovich2010cross, genton2015cross}

The following Lemma is often referred to as Cram\'er's Theorem, which is a multivariate generalization of Bochner's Theorem. 
\begin{lemma}[\cite{cramer1940theory}]\label{lem:cramer}
Let $\widetilde{K}:\RR^p\to \RR^{k\times k}$ be a matrix-valued function with elements $(\widetilde{K}(x))_{ij} = \widetilde{K}_{ij}(x)$ and spectral density $\widetilde{\rho}_{ij}$. Then, $\widetilde{K}$ is positive definite, hence defines a $k$-variate GP on $\RR^p$, if and only if $\widetilde{\rho}(\omega)=\{\widetilde{\rho}(\omega)\}_{i,j=1}^k$ is positive semi-definite for any $\omega\in\RR^p$.
\end{lemma}
We prove the following related result.
\begin{theorem}\label{thm:PDkk}
Let $K:\RR^p\times \mathscr{C}\times\mathscr{C}\to\RR $ be a  function with $K_{ij} = K(\cdot,c_i,c_j)$ being stationary on $\RR^p$ and spectral densities $\rho_{ij}$. Then, $K$ is positive definite, hence defines a semi-stationary GP on $\RR^p\times \mathscr{C}$, if and only if $\rho(\omega)=\{\rho(\omega)\}_{i,j=1}^k$ is positive semi-definite for any $\omega\in\RR^p$.
\end{theorem}
Note that Theorem~\ref{thm:BochnerZ2Z2} is a special case of Theorem~\ref{thm:PDkk} when $k=2$ but can be proved differently. As a result, the connection between Bochner's Theorem on $\RR^p\times\mathscr{C}$ and Cram\'er's Theorem on $\RR^p$ is analogous to the relationship between multi-group GPs and multi-variate GPs.

\subsection{Special cases of the MGGP}\label{sec:special_cases}
We discuss the relationship between the MGGP and several commonly used GPs for multi-group data. We find that each of these models is a special case of the MGGP, which demonstrates the generality of our model. 
The first simple model we consider is the separated GP, which models each group separately with
$k$ independent GPs on $\RR^p$. Let $K_1,\cdots,K_k$ be the covariance functions of these $k$ GPs.
\begin{proposition}
The SGP described above is equivalent to MGGP with covariance function 
\[
K((x,c_i),(x',c_j))=\begin{cases}
K_i(x,x') & i=j\\
0 & i\neq j.
\end{cases}
\]
That is, the SGP is a special case of the MGGP with zero between-group correlation. 
\end{proposition}

Another simple model ignores the group structure entirely, Thus, we treat each observation $(x,c)$ as simply $x$ by dropping the label $c\in\mathscr{C}$. Then, we can consider a GP on the union of all observations across all groups. Assume the covariance function of the UGP is $K_\cup$, which is a GP over $\RR^p$. We obtain the following.
\begin{proposition}
$K_\cup$ is equivalent to MGGP with covariance function
\[
K((x,c_i),(x',c_j))=K_\cup(x,x').
\]
That is, the within group and between-group correlations are all equal. 
\end{proposition}

Finally, we consider the hierarchical GP:
\[
\mu_i\sim GP(0,K_g)\;;\quad Z\given c_i\sim GP(\mu_i, K_z).
\]
where $K_g$ is the covariance function at the group level, while $K_z$ is the covariance function at the covariate level. 
\begin{proposition}
The HGP is equivalent to MGGP with covariance function
\[
K((x,c_i),(x',c_j))=K_g(x,x')+\mathrm{1}_{\{i=j\}}K_z(x,x').
\]
That is, the HGP is a special case of the MGGP with identical within-group dependencies and identical across-group dependencies. 
\end{proposition}

These results reveal that the MGGP is a non-trivial generalization of existing GPs that allows heterogeneity through a variety of flexible covariance functions that it accommodates.

\section{Experiments} \label{sec:experiments}
\subsection{Simulations}
\subsubsection{Demonstrating the MGGP's relationship to related models}
We designed and conducted an experiment to ascertain the MGGP's ability to recover the Separated GP, Union GP, and Hierarchical GP as special cases. We generated synthetic data from each of these models using the likelihood function corresponding to Equation \eqref{eq: mggp_regression} using $k=2$ groups. We specified a zero mean, i.e., $\mu(x; c_j)=0$ for both groups, and specified the latent process through the covariance function corresponding to the three models. We set $b=\sigma^2=a=1$ in
\begin{equation}\label{eq:multigroup_RBF_experiments}
K((x,c_i),(x',c_j))=\frac{\sigma^2}{(a^2d_{ij}^2+1)^{p/2}}\exp\left\{-\frac{b^2\|x-x'\|^2}{a^2d_{ij}^2+1}\right\}.
\end{equation}
(Note that $a$ is only used in the generation of data from the MGGP). We also assumed $\tau_1^2 = \tau_2^2 = \tau^2$ in Equation \eqref{eq: mggp_regression} and used $\tau^2 = 0.1$ to generate our data.
Using these settings, we generated $n_1=n_2=100$ measurements for each group.

We computed the $\log$ marginal likelihood of the data, i.e., $N(y \given 0, K_{\theta} + D_{\tau})$ under each model for each data set. For the SGP, UGP, and HGP, we used the RBF covariance function, $K(x,x')=\sigma^2 \exp\{-b^2\|x-x'\|^2\}$.
For the MGGP, we used the ``multi-group'' version of the RBF (Equation \eqref{eq:multigroup_RBF_experiments}). When computing the likelihood under each model, we fix $b, \sigma^2$, and $\tau^2$ to their true values, for $a$ we use a grid of values, $a = 10^{-5}, 10^{-4}, \dots, 10^{2}$ and specified $D_{\tau} = \tau^2I_{n}$ 

\begin{figure}
    \centering
    \includegraphics[width=\textwidth]{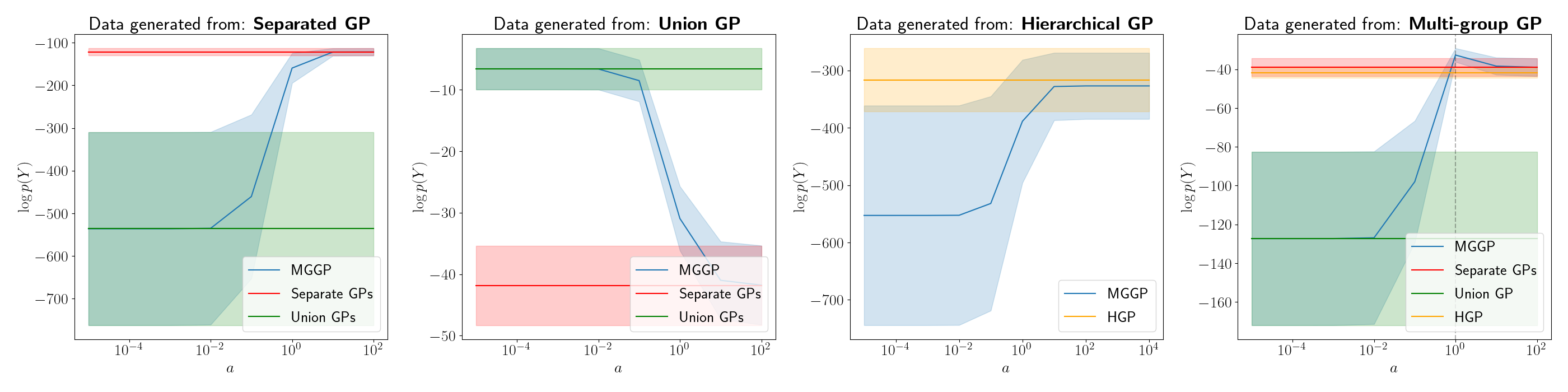}
    \caption{\textbf{Comparison between the \emph{MGGP}, \emph{Separated GP}, \emph{Union GP}, and \emph{Hierarchical GP}}. Using two-group data generated from each of the four models, we computed the log marginal likelihood of the data under each model. For the MGGP, we used the covariance function in Equation \eqref{eqn:RBF} and used a range of different values the parameter $a$. In the rightmost plot, the dashed vertical line indicates the true value of $a$ used for data generation. We use an RBF covariance function for the \emph{Separated GP} and \emph{Union GP}, which does not have an $a$ parameter. We repeated this experiment $20$ times, and the bands in each plot represent $95\%$ confidence intervals.}
    \label{fig:simulation_comparison}
\end{figure}

We find that the MGGP performs at par with the UGP, SGP, and HGP in the expected regimes (\autoref{fig:simulation_comparison}). In particular, the MGGP recovers the performance (as measured by the $\log$ marginal likelihood) of the SGP when $a$ is large, and it recovers the performance of the UGP as $a \rightarrow 0$. For data generated from the MGGP, we find that the likelihood peaks at the true value of $a$ and is higher than all other models at this value. These results i) serve as a demonstration of the role of the $a$ parameter; ii) confirm numerically that the MGGP recovers these models in certain regimes; and iii) suggest that the MGGP is a viable generalization of these models.

\subsubsection{Estimation and inference for the MGGP}
\label{subsec:estimation_and_inference}
In our previous experiment, we used the multi-group covariance function in Equation \eqref{eqn:RBF} and fixed the value of $a$. However, in practice we will need to estimate this and all other covariance parameters from the data. Thus, we next evaluate our ability to fit the parameters of the MGGP using both maximum likelihood estimation and fully Bayesian posterior inference.

\begin{figure}
    \centering
    \includegraphics[width=\textwidth]{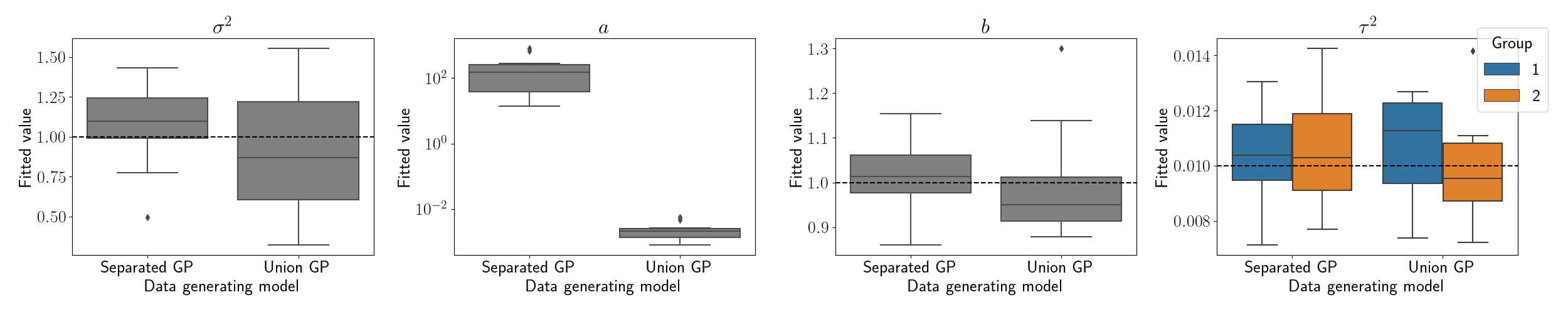}
    \caption{\textbf{Covariance function parameter estimation}. Using data generated from the SGP and UGP, we fit the MGGP by finding the maximum likelihood estimate for the true parameters of the kernel function in Equation \eqref{eqn:RBF}.}
    \label{fig:alpha_optimized}
\end{figure}

\paragraph{Maximum likelihood estimation.}
We first conducted an experiment where we generated data from the SGP and UGP models as in the previous section.
We maximize the collapsed or marginalized likelihood corresponding to Equation~\eqref{eq: mggp_regression}, i.e., $\mathcal{N}(y\given 0, K_{\theta} + \tau^2I_{n})$
with respect to $\theta = \{a,b,\sigma^2\}$ and a common measurement error variance $\tau^2$, where $\theta$ corresponds to the three parameters in the multi-group RBF covariance function in Equation \eqref{eqn:RBF}. 
We used a conjugate gradient ascent algorithm \citep{nocedal2006numerical} to obtain the joint estimates of $\{\theta, \tau^2\}$.
We implemented the algorithm in Python using the JAX software framework~\citep{jax2018github}, which is designed for fast computation, compilation, and automatic differentiation. Experiments were run on an internal computing cluster using a 320 NVIDIA P100 GPU. 
We found that the MLE for $a$ was consistently high for data generated from the SGP and low for the UGP data, as expected (\autoref{fig:alpha_optimized}, middle panel). Additionally, the estimation was able to recover the true values for $\sigma^2$ and $b$ (\autoref{fig:alpha_optimized}, left and right panels).

\begin{figure}
    \centering
    \includegraphics[width=0.4\textwidth]{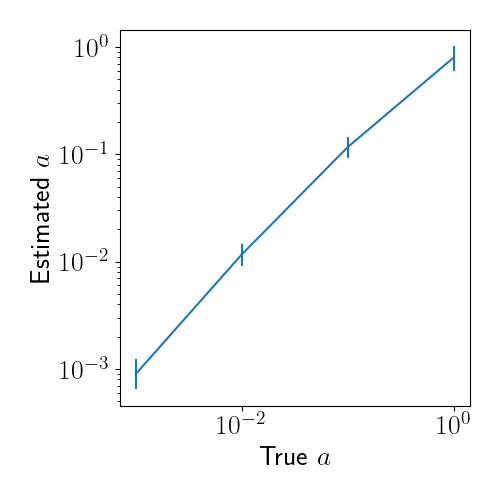}
    \caption{\textbf{Recovering $a$ with ML estimation}. We generated synthetic data from the MGGP at different values for $a$ and fit the MGGP to these data sets. Here, we fix all other parameters to their true values. We find that we are able to recover the true value of $a$ to some extent.}
    \label{fig:alpha_recovered}
\end{figure}

Next, we evaluated whether the MGGP could recover the true value of $a$ from synthetic data sampled from the MGGP model. Here, we generated four datasets, each with a different value of $a$ for $a \in \{10^{-3}, 10^{-2}, 10^{-1}, 10^{0}\}$. We again optimized all parameters jointly by maximizing the marginal likelihood, and we examined the estimated value of $a$ for each. We repeated this experiment ten times.
We found that we could consistently estimate a reasonable value of $a$ (\autoref{fig:alpha_recovered}). While the estimated values do not exactly coincide with the true values, they still showed a monotone relationship. Together, these results show that likelihood-based parameter estimation is feasible in the MGGP model and that existing estimation and optimization techniques can be successfully applied.

\paragraph{Bayesian Inference.}
Next, we used a simulated dataset generated in the same manner as above to 
we conduct a full Bayesian analysis for the MGGP model as described in Section~\ref{sec: mggp_model}. Here, we also include a group-specific intercept for the $k=2$ groups given by $\beta = (\beta_1, \beta_2)^{\top}$, with the data generated according to $\beta_1=1, \beta_2=2$. We form the $(n_1 + n_2) \times 2$ binary design matrix $F$ appropriately in order to apply the group-specific intercept in the model.
With $\theta = \{a,b,\sigma^2\}$
the prior distribution in Equation \eqref{eq: mggp_collapsed} is specified as
\[
p(\theta, \tau^2, \beta) = p(a, b, \sigma^2, \tau^2, \beta) = IG(a \given \alpha_{a}, \alpha'_{a}) IG(b \given \alpha_{b}, \alpha'_{b}) IG(\sigma^2\given \alpha_{\sigma}, \alpha'_{\sigma}) IG(\tau^2 \given \alpha_{\tau}, \alpha'_{\tau}) N(\mu_{\beta}, V_{\beta})\;,
\]
where we set $\alpha_a=\alpha'_a=\alpha_b = \alpha'_b=\alpha_{\sigma} = \alpha'_{\sigma} =\alpha_{\tau} =\alpha'_{\tau} = 1$, $\mu_{\beta} = 0$ and $V_{\beta}^{-1} = \mathrm{I}$. We sample from the posterior distribution in Equation \eqref{eq: mggp_collapsed} using a Hamiltonian No U-Turn Sampling \citep{hoffman2014no} algorithm as implemented in the \texttt{Stan} programming environment \citep{stan2020stan,pystan}. 

\begin{figure}
    \centering
    \includegraphics[width=0.9\textwidth]{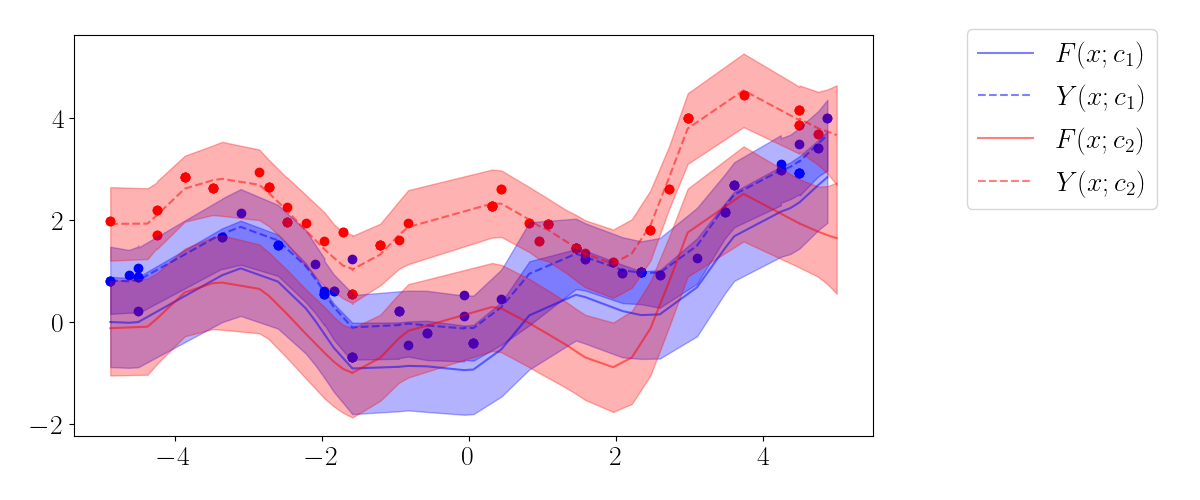}
    \caption{\textbf{Posterior predictive distribution under the MGGP}. The points represent training data; the solid lines show the means of the latent processes $F(x; c)$; the dashed lines represent the predictive means of $Y(x; c)$; and the shaded areas around the lines are twice the standard deviation of the PPD at the corresponding input points.}
    \label{fig:mggp_predictive_samples}
\end{figure}

We ran four chains with dispersed initial values for $400$ iterations each. Convergence was diagnosed after $200$ iterations using visual inspection of autocorrelation plots (\autoref{fig:mggp_autocorrelation}) and computation of Gelman-Rubin R-hat and Monte Carlo standard errors, and the subsequent $800$ samples were retained for posterior inference. The posterior median and $95\%$ credible intervals are presented in \autoref{fig:mggp_hyperparameter_samples}, which shows that the posterior samples of the covariance function parameters center around their true values (\autoref{fig:mggp_hyperparameter_samples}). 
We also sample from the posterior predictive distribution (PPD), $p(Y(x_0; c_j)\given y, F)$ (see Section~\ref{sec: mggp_model}), for a collection of new inputs or test cases. These are presented in \autoref{fig:mggp_predictive_samples}. Notably, because all of the MGGP assumptions are encoded in the covariance function, any appropriate method for estimation and inference in standard GPs can be applied. 

\paragraph{Prediction}
As a final experiment with simulated data, we evaluated the MGGP in its ability to predict held-out values in a Gaussian process regression task. We generated data from a GP regression model (Equation \eqref{eq: mggp_regression}), using the Separated GP, Union GP, HGP, and MGGP. We then fit these models on each of the data sets, using $50\%$ of the data for fitting and computing predictions for the other $50\%$. We use the posterior mean as a point prediction for each of the $n^\star$ held-out samples:
\begin{equation}\label{eq:posterior_predictive_mean}
    \mu^\star = K_{X^\star X} K_{XX}^{-1} y,
\end{equation}
where $K_{X^\star X}$ is the $n^\star \times n$ matrix of covariance function evaluations for each pair of test and train samples, and $K_{XX}$ is the $n \times n$ matrix of covariance function evaluations for each pair of train samples. We mean-center the data for each group. To measure the quality of the predictions, we compute the mean squared error of the predictions,
\begin{equation}\label{eq:MSE}
    E = \frac{1}{n^\star} \sum_{i=1}^{n^\star} (y_i - \mu^\star_i)^2
\end{equation}

\begin{figure}
    \centering
    \includegraphics[width=0.7\textwidth]{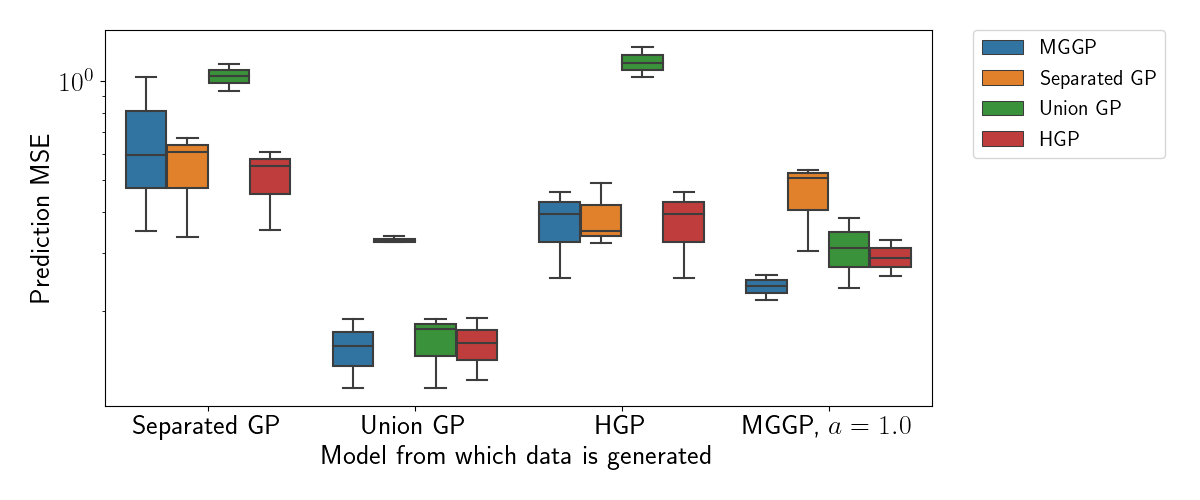}
    \caption{\textbf{GP Prediction with simulated data}. We generate data from each of the four models --- the SGP, UGP, HGP, and MGGP --- and fit each of these models too all datasets. Prediction error was computed on a hold-out dataset.}
    \label{fig:prediction_simulated}
\end{figure}

We found that the MGGP emulates the performance of the SGP, UGP, and HGP on their respective simulated datasets (\autoref{fig:prediction_simulated}), while with data generated from the MGGP itself, the MGGP clearly outperformed the other models.
While the Separated GP can be expected to perform well when each group has a large sample size, a primary benefit of the MGGP is its ability to share information across similar groups when the sample size is limited. Thus, we expect the performance of the MGGP to improve relative to the SGP in settings in which some groups have a small number of samples but are closely related to other, larger groups.

\begin{figure}
    \centering
    \includegraphics[width=0.5\textwidth]{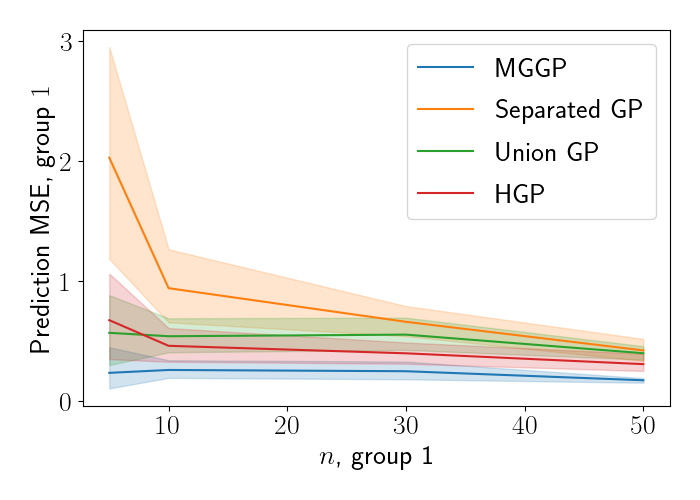}
    \caption{\textbf{Prediction using simulated data with imbalanced groups}. We perform a prediction experiment with $k=3$ groups. To generate a series of datasets, we fix the sample size of groups $c_2$ and $c_3$ to be $50$, and vary the sample size of group $c_1$. }
    \label{fig:prediction_simulated_group_imbalance}
\end{figure}

To test this hypothesis, we conducted another multi-group GP regression experiment in which we sought to predict the held-out values for one group that contained few samples. Specifically, we generated a synthetic dataset containing three groups, where group $1$ and group $2$ are similar to one another, and group $3$ is dissimilar from the other two. We then generated a series of datasets, varying the number of samples in group $1$ for each. Finally, we computed the prediction error for the SGP, UGP, HGP, and MGGP.
We found that the MGGP outperformed the other models, especially when the sample size for group $1$ was small (\autoref{fig:prediction_simulated_group_imbalance}), confirming our hypothesis. This result shows that the MGGP thrives relative to other setups when the sample size for some groups is limited, as it can most effectively leverage information from similar groups.

\subsection{Application to GTEx tissue samples}
We next applied the MGGP to a large gene expression data set collected by the Genotype-Tissue Expression (GTEx) project~\citep{gtex2020gtex}. The GTEx v8 data contain measurements from 17382 samples that span 52 tissue types collected from 838 human donors; see Appendix \ref{appendix:gtex_data} for a full list of tissue types and the sample size for each tissue. Along with gene expression profiling, a variety of additional metadata characteristics were collected, including demographic variables and tissue health measurements.

In these experiments, we use Gaussian process regression models to analyze the relationship between a sample's gene expression profile and its ischemic time (the duration of time between death and tissue collection). Previous work has shown a robust relationship between gene expression and ischemic time~\citep{musella2013effects,ferreira2018effects}; however, whether this relationship exhibits tissue-specific patterns remains largely unknown. In these experiments, the groups correspond to tissue types.

\subsubsection{Exploring group similarities with the MGGP}
As an initial test with the GTEx data, we applied the MGGP to just two tissue types at a time. These experiments aim to validate that the MGGP can appropriately model known associations between similar groups.

\begin{figure}
    \centering
    \includegraphics[width=\textwidth]{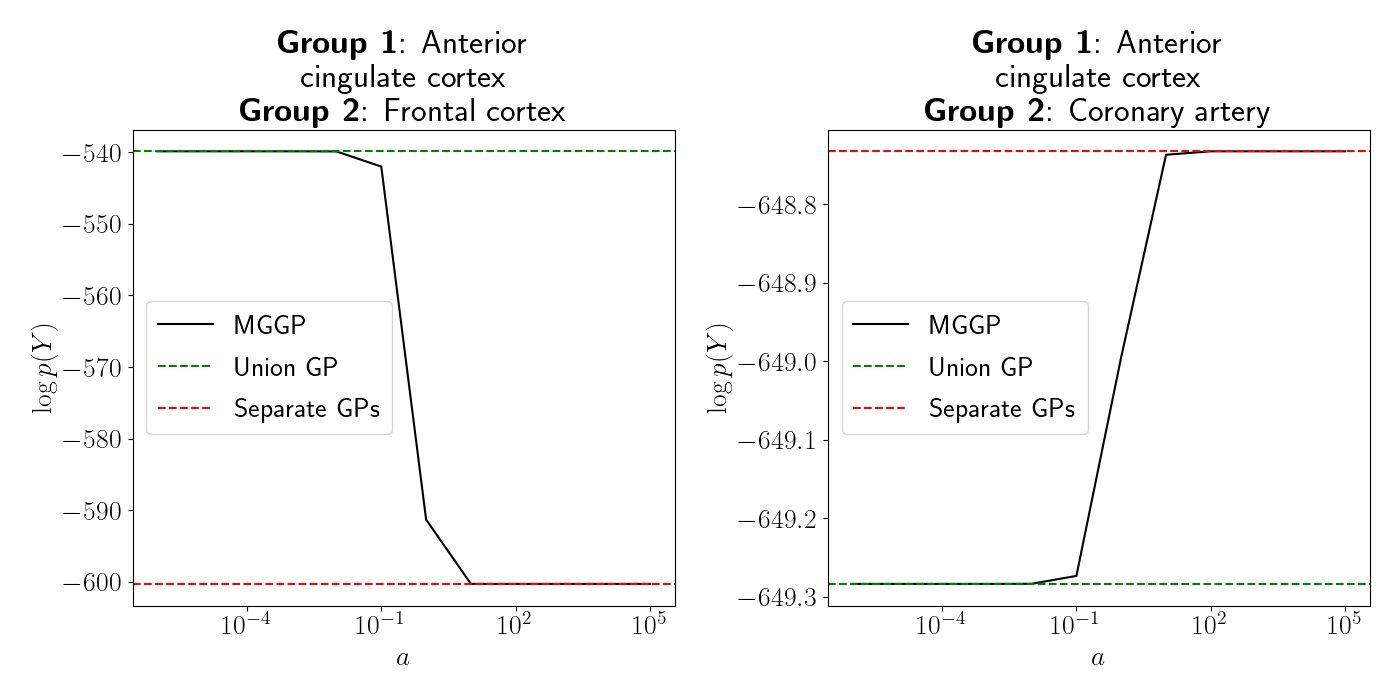}
    \caption{\textbf{Likelihood of GTEx gene expression data under the MGGP}.  For two pairs of tissues, we computed the log marginal likelihood of the data under the MGGP with $a$ set to be a range of values. Similar tissue types (anterior cingulate cortex and frontal cortex) prefer low values of $a$, while more dissimilar tissues (anterior cingulate cortex and coronary artery) prefer high values of $a$.}
    \label{fig:gtex_twogroup_arange}
\end{figure}

\paragraph{Examining model fit via the log marginal likelihood} In a preliminary experiment, we examined three tissue types: tibial artery ($n = 657$), coronary artery ($n = 238$), and breast ($n = 456$). First, for each of the three pairs of tissues, we fit the MGGP with maximum likelihood estimation, as described in Section \ref{subsec:estimation_and_inference}, using the multi-group RBF covariance function in Equation \eqref{eq:multigroup_RBF_experiments}. In this experiment, we fixed $a$ to one value in a preset range, and found the maximum likelihood estimates of the remaining parameters. Using these maximum likelihood estimates and the fixed $a$, we then computed the $\log$ marginal likelihood of the data,
\begin{equation*}
    \log p(y | X, a, \widehat{b}, \widehat{\sigma^2}, \widehat{\tau^2}) = -\frac{k}{2} 2\pi - \frac12 \det (K_{XX} + \widehat{\tau^2} I) -\frac12 y^\top (K_{XX} + \widehat{\tau^2} I)^{-1} y ,
\end{equation*}
where $K_{XX}$ is the $(n_1 + n_2) \times (n_1 + n_2)$ matrix of covariance function evaluations for each pair of samples. We also fit the Separated GP and Union GP for each pair of tissues using the standard RBF kernel, and computed the $\log$ marginal likelihood of the data under these models.
Examining the log marginal likelihood across varying values of $a$ (\autoref{fig:gtex_twogroup_arange}), we found that two tissue types that are expected to be similar to one another --- tibial and coronary artery --- showed a higher marginal likelihood under small values of $a$ ($a \lessapprox 0.01$), while tissues that have unique expression patterns --- tibial artery and breast --- showed a higher marginal likelihood under large values of $a$ ($a \gtrapprox 10$). In both cases, the MGGP gracefully recovered the Separated GP and Union GP marginal likelihoods for $a \to \infty$ and $a \to 0$, respectively. This result implies that the MGGP could be a viable strategy not only for sharing information across groups, but also for learning the group relationships themselves.

\paragraph{Recovering group similarities} Next, we conducted a similar experiment in which we also estimated $a$ from the data itself (along with all other model and covariance function parameters), again using maximum likelihood estimation. In this experiment, we applied the model to all $52$ tissue types.
We fit the MGGP for every pair of tissues, and extracted $\widehat{a}_{MLE}$ for each pair. This experiment yields $\frac12(52 \times 51) = 1326$ estimated values of $a$ (one for each pair of tissue types).

\begin{figure}[!h]
    \centering
    \includegraphics[width=0.9\textwidth]{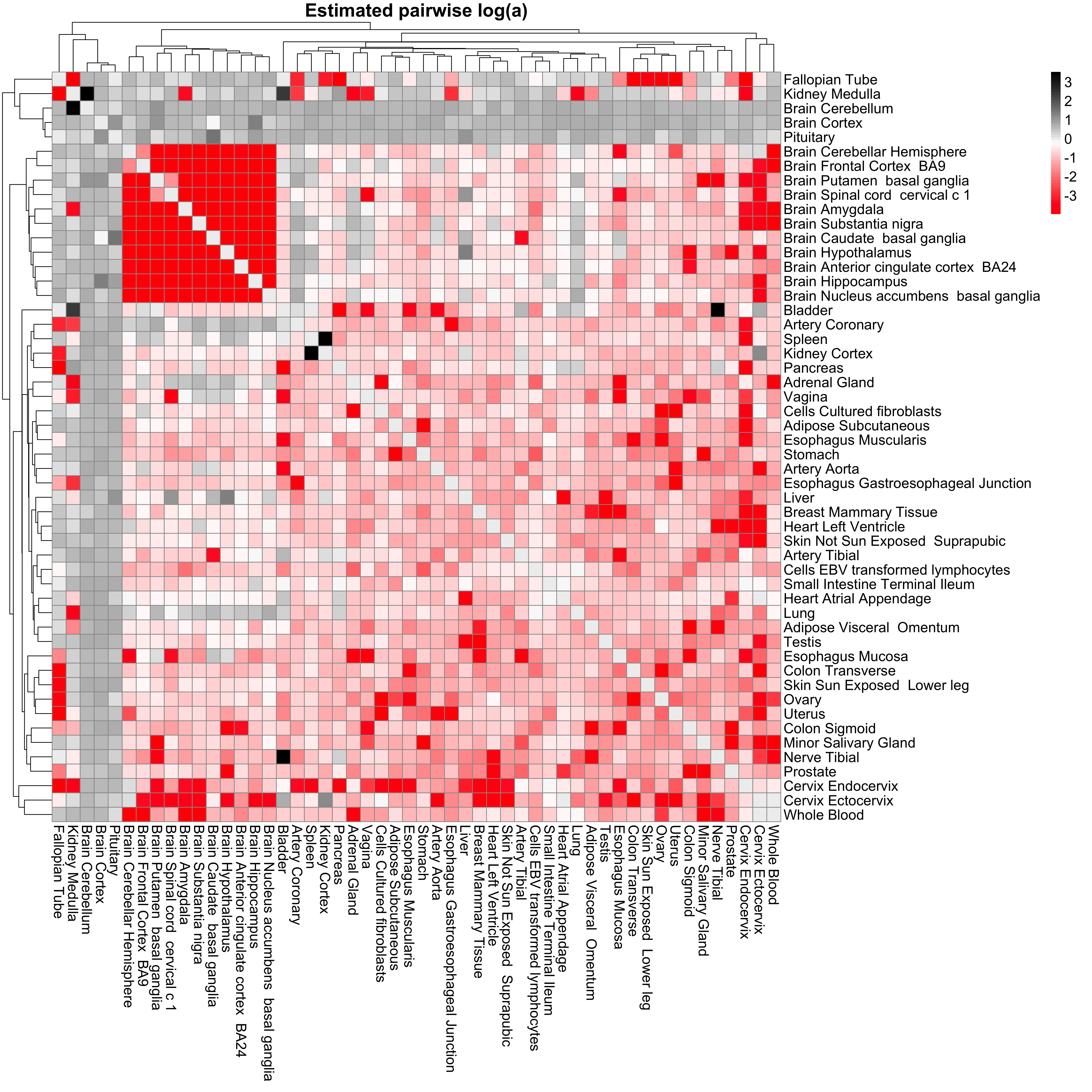}
    \caption{\textbf{Estimation of $a$ for each pair of GTEx tissue types}. Cell $ij$ in the heatmap represents $\log_{10}(a_{ij})$, where $a_{ij}$ is the maximum likelihood estimate of $a$ when fitting the MGGP using tissues $i$ and $j$. Lower values of $a$ (red) indicate higher similarity, while higher values of $a$ (black) indicate lower similarity.}
    \label{fig:gtex_pairwise_a_heatmap_full}
\end{figure}

We found that the estimated values of $a$ reflect many of the expected relationships between the tissue types (\autoref{fig:gtex_pairwise_a_heatmap_full}). Most notably, we found that 11 subtypes of brain tissue yield low values for $a$, suggesting that gene expression changes in these tissue types in a similar manner as ischemic time changes. 

\paragraph{Recovering group similarities with group-specific errors} As a final test of the MGGP in the two-group setting, we performed a similar pairwise analysis as in the previous paragraph, but here we allowed each group to have its own noise variance $\tau_j$ as in \autoref{eq: mggp_regression}. We apply this model to a subset of the tissue types: Brain Cortex, Brain Hypothalamus, Esophagus Gastroesophageal Junction, and Esophagus Mucosa. We fit the MGGP to each of the six pairs of tissues and extracted the estimated value for $a$. We found that the resulting values of $a$ recovered similarities between tissue types: the MGGP obtained a small estimated value for the two brain tissues and for the two esophagus tissues, while all other values were quite large (\autoref{fig:gtex_pairwise_a_heatmap_small_group_specific_variances}). This small-scale experiment demonstrates that the MGGP behaves similarly when group-specific noise variances are included. More generally, these experiments indicate that, in the pairwise case, the estimated value of $a$ reflects prior biological knowledge about similarity of tissue types.

\subsubsection{Leveraging group relationships for prediction}
We next conducted a series of experiments with the GTEx data to test whether the MGGP's model of group relationships could improve downstream predictive performance relative to standard GPs. In each of these experiments, we randomly split the data into 50\% training samples and 50\% testing samples, and mean-centered each dataset within each group. We fit the MGGP on the training set using maximum likelihood estimation, using the multi-group RBF kernel in all experiments here (Equation \eqref{eq:multigroup_RBF_experiments}). Then, we computed the predicted ischemic time for each test sample using the predictive mean (Equation \eqref{eq:posterior_predictive_mean}) for the ischemic time for each gene expression sample and computed the mean squared error relative to the true ischemic time (Equation \eqref{eq:MSE}). For comparison, we also fit the Union and Separated GPs using the standard RBF covariance function and the Hierarchical GP using the standard RBF as both the within- and between-group covariance functions.

\paragraph{Benchmarking the MGGP, UGP, SGP, and HGP} We first performed a prediction exercise using eight GTEx tissue types: Anterior Cingulate Cortex, Frontal Cortex, Cortex, Breast, Tibial Artery, Coronary Artery, Uterus, and Vagina. Given \emph{a priori} similarities between tissue types from different body parts, we specify the ``distance'' between two similar tissue types ($d_{ij}$) to be small, while the distance between dissimilar tissue types is large. Here, we identify sets of similar tissues as \{Anterior Cingulate Cortex, Frontal Cortex, Cortex\}, \{Breast\}, \{Tibial Artery, Coronary Artery\}, and \{Uterus, Vagina\}. For each pair of tissues within each of these subsets, we specify the distance to be $d_{ij} = 10$, and for each pair of tissues between subsets, we specify the distance as $d_{ij} = 0.1$. The distance from a group to itself is zero, $d_{ij} = 0$ for $i=j$. Note that, for the multi-group RBF covariance function (Equation \eqref{eq:multigroup_RBF_experiments}), only the \emph{relative} distances between groups are relevant, as the $a$ parameter as a scaling factor on these relative distances, and we estimate $a$ from the data. Thus, our distance settings imply an assumption that tissue types within each of these subsets are ten-fold more similar than tissues between the subsets.

\begin{figure}
    \centering
    \includegraphics[width=0.9\textwidth]{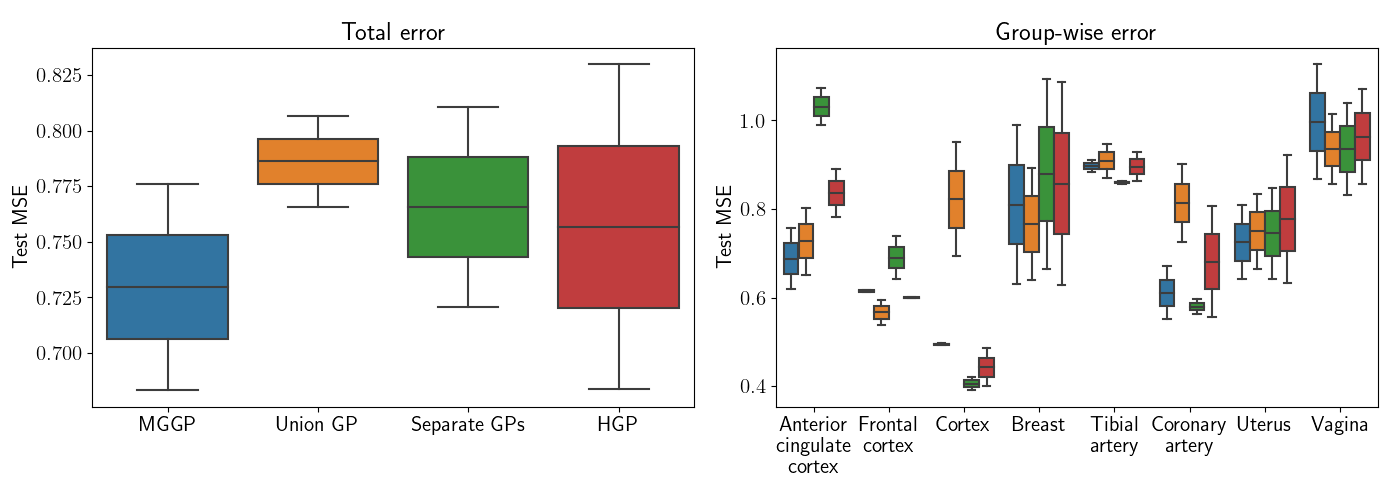}
    \caption{\textbf{Prediction with GTEx data}. We fit each of the four models in a regression setting, where the covariates are gene expression profiles of samples from various tissue types, and the response is each sample's ischemic time. We compute the error on a held-out set of data.}
    \label{fig:gtex_prediction}
\end{figure}

Examining the prediction error for each of the four models, we find that MGGP outperforms UGP, SGP and HGP in terms of overall prediction error, although the HGP performs somewhat comparably (\autoref{fig:gtex_prediction}, left panel). Further examining the error for each tissue, we find that the performance varies substantially across tissue types (\autoref{fig:gtex_prediction}, right panel). 

\paragraph{Fine-tuning group distances} While we observe that the MGGP performs well in a prediction setting, the \emph{a priori} distances between tissues were chosen somewhat heuristically as a coarse approximation of prior biological knowledge. However, one may wonder whether these distances can be chosen more precisely, or perhaps even be learned from the data itself. To test this, we performed another prediction experiment with three tissue types: Anterior Cingulate Cortex, Frontal Cortex, and Breast. Instead of hard-coding the distances, we used a data-driven approach that uses two steps. In the first step, we fit a two-group MGGP on each of the three pairs of tissues, and used the resulting MLE for $a$ as the distance between two tissues, $d_{ij} = \widehat{a}_{ij}$. For the two cortex tissues, we find $a \approx 10^{-4}$, and and for each of the cortex and breast pairs, we find $a \approx 1$. In the second step, we fit the full three-group MGGP using these learned distances. We also fit the HGP for comparison. 

\begin{figure}
    \centering
    \includegraphics[width=0.9\textwidth]{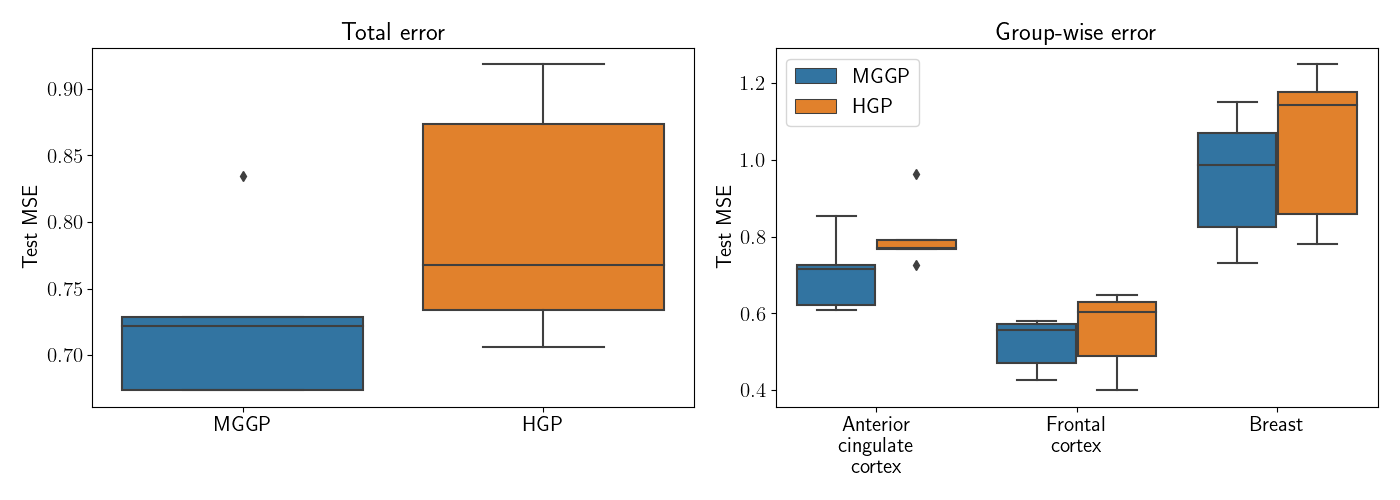}
    \caption{\textbf{Fine-tuning group distances and comparison of the MGGP and HGP via prediction with GTEx data}.We use just three tissue types and carefully tune the pairwise distances between the groups using a data-driven approach.}
    \label{fig:gtex_prediction_mggp_vs_hgp}
\end{figure}

We find that after using this data-driven approach to select the group-wise distances, the MGGP performs better than before, both in terms of overall prediction error and group-wise error (\autoref{fig:gtex_prediction_mggp_vs_hgp}). Moreover, we find that the MGGP more clearly outperforms the HGP in this case. This difference is likely due to the nature of the HGP's model for group relationships: the HGP treats all groups symmetrically, in the sense that the only relevant quantity is whether two samples are in the same group or different groups. In contrast, the MGGP relies on the specific group identities of any two samples, as well as the distance between those groups. This result implies that carefully defining the group relationships can be important for the MGGP, and that the demonstrated data-driven approach is a viable strategy for doing so.

\paragraph{Assessing univariate relationships} The MGGP regression framework also allows for assessing relationships between the outcome variable and individual explanatory variables. To demonstrate this, we again leveraged data for three GTEx tissue types: Anterior Cingulate Cortex, Frontal Cortex, and Breast. In this experiment, we fit the MGGP to these data using just one gene at a time, finding a one-dimensional regression function describing each gene's relationship to ischemic time. We fit the model using maximum likelihood estimation and the multi-group RBF covariance function (Equation \eqref{eq:multigroup_RBF_experiments}), and we used the between-group distances found in the previous experiment. We then computed the predicted mean (Equation \eqref{eq:posterior_predictive_mean}) for a dense grid of input points, and we examined the resulting regression functions. 

\begin{figure}
    \centering
    \includegraphics[width=0.6\textwidth]{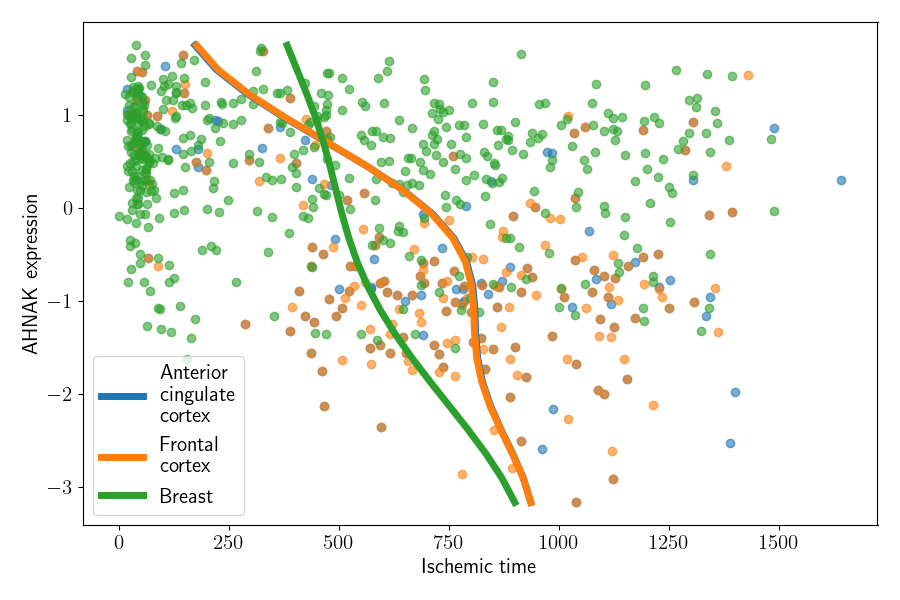}
    \caption{\textbf{Examining univariate relationships with the MGGP}. For three tissue types, each sample's expression of the gene \emph{AHNAK} (x-axis) and its ischemic time (y-axis) are shown by the scatter points. The solid lines show the MGGP's posterior predictive mean when fit on this single gene in these three tissues.}
    \label{fig:gtex_one_gene_relationship}
\end{figure}

We find that the MGGP identifies several trends in the relationship between gene expression and ischemic time that show group-level trends. For instance, we find that expression of the gene \emph{AHNAK}, which codes for a neuroblast differentiation-associated protein, decays similarly for the two cortex tissues, while this decay pattern follows a different trend in breast tissue (\autoref{fig:gtex_one_gene_relationship}). Moreover, we find that this relationship is robust to the choice of group distances (\autoref{fig:gtex_one_gene_relationship_equidistant_ENSG00000124942}). This experiment demonstrates that the MGGP can be used for more specifically characterizing the group-wise relationships present in a dataset.

\section{Discussion}
\label{sec:discussion}
In this work, we present the multi-group Gaussian process (MGGP), a flexible approach for modeling complex dependencies in datasets with subgroup structure. We present several options for constructing valid covariance functions on $\RR^p\times\mathscr{C}$, and we show that this structure generalizes several existing GP models. We demonstrate the behavior of the MGGP through a series of simulation experiments and an application to a dataset of gene expression measurements.

Several future directions remain to be explored. First, this paper lays the groundwork for developing new positive definite covariance functions on $\RR^p\times\mathscr{C}$. A particularly interesting direction is to construct covariance functions whose within-group and between-group correlations exhibit fundamentally different structure (e.g., the within-group correlation may be Matérn-like, while the between group correlation may be RBF-like). Second, as briefly mentioned in Section~\ref{sec: mggp_model}, recent advances in scalable GPs could be applied to the MGGP. For example, sparsity-inducing GPs have received much attention in the recent past 
\citep[see, e.g.,][]{datta2016hierarchical, katzfuss2021general, pbf2021} 
and such methods can be applied to the class of MGGP models presented here. Finally, although we primarily focused on GP regression in this paper, the MGGP can be applied in any GP-based model. Thus, there could be benefits from using the MGGP in classification models, latent variable models, and other applications. We envision the MGGP being a flexible tool in a variety of contexts.

\section{Appendix}
\subsection{Proof of Proposition \ref{prop:PD_cat}}
\begin{proof}
First, assume $K_2$ is positive definite, and let the observation set be $\mathscr{C}=\{c_1,\cdots,c_k\}$. Then the covariance matrix is exactly $C$, which is positive definite as well.

We assume $C$ is positive definite. Given any observation set, we can rearrange the observations to be
$\{\underbrace{c_1,\cdots,c_1}_{n_1},\cdots,\underbrace{c_k,\cdots,c_k}_{n_k}\}$. Then the corresponding covariance matrix $\Sigma$ can be written as the following block matrix:
$$\Sigma=  \begin{bmatrix} C_{11}\bm{1}_{n_1\times n_1} & C_{12}\bm{1}_{n_1\times n_2}&\cdots     &C_{1k}\bm{1}_{n_1\times n_k}\\
        C_{21}\bm{1}_{n_2\times n_1} & C_{22}\bm{1}_{n_2\times n_2} &\cdots & C_{2k}\bm{1}_{n_2\times n_k}\\
        \vdots & \ddots& & \vdots \\
        C_{k1}\bm{1}_{n_k\times n_1} & C_{k2}\bm{1}_{n_k\times n_2} &\cdots &C_{kk}\bm{1}_{n_k\times n_k}
    \end{bmatrix},$$
    where $\bm{1}$ is the matrix with all entries equal to $1$. Let $P_i\in\RR^{n_i\times n_i}$ be rotation matrices such that $P_i\bm{1}_{n_i}=\sqrt{n_i}\bm{e}_{n_i}$, where $e_{n_i}=[1,0,\cdots,0]^\top\in\RR^{n_i}$, $i=1,\cdots,k$, and let $P\coloneqq\diag\{P_1,\cdots,P_k\}\in \RR^{\sum_{i}n_i\times \sum_i n_i}$. Then,
\begin{align*}
    P\Sigma P^\top& = 
    \begin{bmatrix} 
        P_1 & &\\ & \ddots &\\ &&P_k
    \end{bmatrix}
    \begin{bmatrix} C_{11}\bm{1}_{n_1\times n_1} & C_{12}\bm{1}_{n_1\times n_2}&\cdots     &C_{1k}\bm{1}_{n_1\times n_k}\\
        C_{21}\bm{1}_{n_2\times n_1} & C_{22}\bm{1}_{n_2\times n_2} &\cdots & C_{2k}\bm{1}_{n_2\times n_k}\\
        \vdots & \ddots& & \vdots \\
        C_{k1}\bm{1}_{n_k\times n_1} & C_{k2}\bm{1}_{n_k\times n_2} &\cdots &C_{kk}\bm{1}_{n_k\times n_k}
    \end{bmatrix}
    \begin{bmatrix}
        P_1^\top & &\\ & \ddots &\\ &&P_k^\top
    \end{bmatrix}\\
    & = 
    \begin{bmatrix}
        C_{11}P_1\bm{1}_{n_1\times n_1} P_1^\top & \cdots & C_{1k}P_1\bm{1}_{n_1\times n_k} P_k^\top \\
        \vdots &\ddots &\vdots\\
        C_{k1}P_k\bm{1}_{n_k\times n_1} P_1^\top &\cdots & C_{kk}P_p\bm{1}_{n_k\times n_k} P_k^\top 
    \end{bmatrix}\\
    & = \begin{bmatrix}
        C_{11}(P_1\bm{1}_{n_1})  (P_1\bm{1}_{n_1})^\top & \cdots & C_{1k}(P_1\bm{1}_{n_1})  (P_k\bm{1}_{n_k})^\top \\
        \vdots &\ddots &\vdots\\
        C_{k1}(P_k\bm{1}_{n_k})  (P_1\bm{1}_{n_1})^\top &\cdots & C_{kk}(P_k\bm{1}_{n_k})  (P_k\bm{1}_{n_k})^\top
    \end{bmatrix}\\
    &= \begin{bmatrix}
        C_{11}n_1\bm{e}_{n_1}\bm{e}_{n_1}^\top  & \cdots & C_{1k}\sqrt{n_1n_k}\bm{e}_{n_1}\bm{e}_{n_k}^\top \\
        \vdots &\ddots &\vdots\\
        C_{k1}\sqrt{n_kn_1}\bm{e}_{n_k}\bm{e}_{n_1}^\top  &\cdots &C_{kk}n_k\bm{e}_{n_k}\bm{e}_{n_k}^\top
    \end{bmatrix}
\end{align*}
As a result, eigenvalues of $\Sigma$ are the same as eigenvalues of $\widetilde{\Sigma} = \begin{bmatrix} C_{11}n_1 & \cdots & C_{1k}\sqrt{n_1n_k}\\
 \vdots &\ddots &\vdots\\
 C_{k1}\sqrt{n_kn_1} &\cdots & C_{kk}n_k
\end{bmatrix}$. Let $v=[\sqrt{n_1},\cdots,\sqrt{n_k}]^\top$ then
$$\widetilde{\Sigma} = C \odot vv^\top.$$
Then by assumption and Schur product theorem \citep{styan1973hadamard}, $\widetilde{\Sigma}$ is positive semi-definite, which finishes the proof.
\end{proof}
\subsection{Proof of Corollary \ref{cly:PDZ2}}
\begin{proof}
By Proposition \ref{prop:PD_cat}, $K_2$ is positive definite if and only if $C$ is positive definite. Under the assumption of homogeneity, $C=\begin{bmatrix}1 & b &\cdots & b\\
b & 1 & \cdots & b\\
\vdots &\vdots &\ddots  &\vdots\\
b & b & \cdots & 1\end{bmatrix}$. 
We can rewrite $C = (1-b)\mathrm{I}_k+b \bm{1}_{k\times k}$. 
Recall that $\bm{1}_{k\times k}$ has eigenvalue $p$ with multiplicity $1$, and the corresponding eigenvector is $\bm{1}_k$. Moreover, $0$ is another eigenvalue with multiplicity $k-1$. As a result, eigenvalues of $\Sigma$ are $1-b+bk$ with multiplicity $1$ and $1-b$ with multiplicity $-1$. We can conclude that $C$ is positive definite if and only if $-\frac{1}{k-1}\leq b \leq 1$. 
\end{proof}

\subsection{Proof of Theorem \ref{thm:st}}
\begin{proof}
By the isometric embedding, finding $K$ is equivalent to finding a semi-isotropic covariance function $K_0$ in $\RR^p\times\RR^{p'}$. For any completely monotone function $\varphi: \RR_+\to \RR$ and positive function $\psi:\RR_+\to \RR_+$ with a completely monotone derivative, \cite{gneiting2002nonseparable} proved that $K_0((x_1,x_1'),(x_2,x_2'))=\frac{\sigma^2}{\left(\psi(\|x_1'-x_2'\|^2)\right)^{p/2}}\varphi\left(\frac{\|x_1-x_2\|^2}{\psi(\|x_1'-x_2'\|^2)}\right)$ is positive definite for any $\sigma^2>0$. As a result,
\[
K((x,c_i),(x',c_j))\coloneqq K_0((x,\iota(c_i)),(x',\iota(c_j)))
\]
is positive definite. 
\end{proof}

\subsection{Proof of Theorem \ref{thm:BochnerZ2}}
Recall the general form of Bochner's Theorem for a locally compact Abelian group:
\begin{lemma}[Bochner's Theorem]
 Let $G$ be a locally compact Abelian group and $\widehat{G}$ be its dual group, then for any continuous positive-definite function $K$ on $G$, there exists a unique positive measure $\mu$ on $\widehat{G}$  such that
 \[
 K(g) = \int_{\widehat{G}}\xi(g)d\mu(\xi).
 \]
\end{lemma}
Note that $G=\RR^p\times \ZZ_2$ is a locally compact Abelian group and the dual group is $\widehat{G}=\RR^p\times U_2$, where $U_2$ is the group of second roots of unity, that is, $U_2 = \{1,-1\}$. $\widehat{G}$ acts on $G$ as
$$(\omega,z)((x,l))=e^{-2\pi i\omega x}z^l.$$

The spectral measure of $K$, denoted by $\mu$ on $\RR^p\times U_2$ splits as $\mu_1\times \mu_2$ where $\mu_1$ is a measure on $\RR^p$ and $\mu_2$ is a measure on $U_2$. Then we claim that
\[
K(x,l) = \sum_{z\in U_2}\int_{\RR^p}e^{-2\pi i\omega x}z^l\rho(\omega,z)d\omega,
\]
where $\rho(\omega,1)=\frac{1}{2}\rho_w(\omega)+\frac{1}{2}\rho_c(\omega)$ and $\rho(\omega,-1)=\frac{1}{2}\rho_w(\omega)-\frac{1}{2}\rho_c(\omega)$.
We derive from the right hand side to the left hand side. First consider the case when $l=0$:
\begin{align*}
  &\sum_{z\in U_2}\int_{\RR^p}e^{-2\pi i\omega x}z^0\rho(\omega,z)d\omega\\
  &= \frac{1}{2}\int_{\RR^p} e^{-2\pi i \omega x}(\rho_w(\omega)+\rho_c(\omega))d\omega+\frac{1}{2}\int_{\RR^p} e^{-2\pi i \omega x}(\rho_w(\omega)-\rho_c(\omega))d\omega\\
  &= \int_{\RR^p} e^{-2\pi i \omega x}\rho_w(\omega)d\omega\\
  & =K_w(x)=K(x,0).
\end{align*}
Similarly, when $l=1$,
\begin{align*}
  &\sum_{z\in U_2}\int_{\RR^p}e^{-2\pi i\omega x}z^1\rho(\omega,z)d\omega\\
  &= \frac{1}{2}\int_{\RR^p} e^{-2\pi i \omega x}(\rho_w(\omega)+\rho_c(\omega))d\omega-\frac{1}{2}\int_{\RR^p} e^{-2\pi i \omega x}(\rho_w(\omega)-\rho_c(\omega))d\omega\\
  &= \int_{\RR^p} e^{-2\pi i \omega x}\rho_c(\omega)d\omega\\
  & =K_c(x)=K(x,1).
\end{align*}

As a result, $K$ is PD $\Longleftrightarrow \rho$ is a positive measure $\Longleftrightarrow \rho_w\geq \rho_c$, and the Theorem follows.

\subsection{Proof of Theorem \ref{thm:BochnerZ2Z2}}

The generalized spectral measure of $K$, denoted by $\mu$ (with density $\rho$) on $\RR^p\times U_2\times U_2$, splits as $\mu_1\times \mu_2$ where $\mu_1$ is a measure on $\RR^p$ and $\mu_2$ is a positive semi-definite measure on $U_2\times U_2$. Then
\[
K(x,l,l') = \sum_{z\in U_2}\sum_{z'\in U_2}\int_{\RR^p}e^{-2\pi i\omega x}z^l\overline{z'^{l'}}\rho(\omega,z,z')d\omega.
\]
We claim that
$\rho(\omega,z,z')=\begin{cases}
\frac{1}{4}\left(\rho_0(\omega)+2\rho_c(\omega)+\rho_0(\omega)\right) & z=z'=1\\
\frac{1}{4}\left(\rho_0(\omega)-\rho_1(\omega)\right) & zz'=-1\\
\frac{1}{4}\left(\rho_0(\omega)-2\rho_c(\omega)+\rho_1(\omega)\right)& z=z'=-1.
\end{cases}$.

We derive from the right hand side to the left hand side. First consider the case when $l=l'=0$:
\begin{align*}
  &\sum_{z\in U_2}\sum_{z'\in U_2}\int_{\RR^p}e^{-2\pi i\omega x}z^0\overline{z'^0}\rho(\omega,z,z')d\omega\\
  &= \frac{1}{4}\int_{\RR^p} e^{-2\pi i \omega x}\left(\rho_0(\omega)+2\rho_c(\omega)+\rho_1(\omega)+\rho_0(\omega)-\rho_1(\omega)+\rho_0(\omega)-\rho_1(\omega)+\rho_0(\omega)-2\rho_c(\omega)+\rho_1(\omega)\right)d\omega\\
  &= \int_{\RR^p} e^{-2\pi i \omega x}\rho_0(\omega)d\omega\\
  & =K_0(x)=K(x,0,0).
\end{align*}
Similarly, when $l=0,l'=1$,
\begin{align*}
   &\sum_{z\in U_2}\sum_{z'\in U_2}\int_{\RR^p}e^{-2\pi i\omega x}z^0\overline{z'^1}\rho(\omega,z,z')d\omega\\
  &= \frac{1}{4}\int_{\RR^p} e^{-2\pi i \omega x}\left(\rho_0(\omega)+2\rho_c(\omega)+\rho_1(\omega)-\rho_0(\omega)+\rho_1(\omega)+\rho_0(\omega)-\rho_1(\omega)-\rho_0(\omega)+2\rho_c(\omega)-\rho_1(\omega)\right)d\omega\\
  &= \int_{\RR^p} e^{-2\pi i \omega x}\rho_c(\omega)d\omega\\
  & =K_c(x)=K(x,0,1).
\end{align*}
When $l=l'=1$,
\begin{align*}
   &\sum_{z\in U_2}\sum_{z'\in U_2}\int_{\RR^p}e^{-2\pi i\omega x}z^1\overline{z'^1}\rho(\omega,z,z')d\omega\\
  &= \frac{1}{4}\int_{\RR^p} e^{-2\pi i \omega x}\left(\rho_0(\omega)+2\rho_c(\omega)+\rho_1(\omega)-\rho_0(\omega)+\rho_1(\omega)-\rho_0(\omega)+\rho_1(\omega)+\rho_0(\omega)-2\rho_c(\omega)+\rho_1(\omega)\right)d\omega\\
  &= \int_{\RR^p} e^{-2\pi i \omega x}\rho_1(\omega)d\omega\\
  & =K_1(x)=K(x,1,1).
\end{align*}
As a result, $K$ is PD $\Longleftrightarrow \rho$ is a positive semi-definite measure $\Longleftrightarrow (\rho_0+2\rho_c+\rho_1)(\rho_0-2\rho_c+\rho_1)-(\rho_0-\rho_1)^2=4(\rho_0\rho_1-\rho_c^2)\geq 0$, and the Theorem follows.

\subsection{Proof of Theorem \ref{thm:MGGPkGP}}

\begin{proof}
Given a Gaussian random field $Z$ on $\mathcal{Y}\times \mathscr{C}$ with covariance function $K$, we prove that $\widetilde{Z}\coloneqq\Phi(Z)$ is a Gaussian $k$-variate random field on $\mathcal{Y}$ with cross-covariance function $\widetilde{K}$. It suffices to check that $\cov(\widetilde{Z}(x),\widetilde{Z}(x'))=\widetilde{K}(x,x')$, for any $x,x'\in\mathcal{Y}$.
\begin{align*}
    \cov(\widetilde{Z}(x),\widetilde{Z}(x'))& = \cov([Z(x,c_1),\cdots,Z(x,c_k)]^\top,[Z(x',c_1),\cdots,Z(x',c_k)]^\top)\\
    & = \begin{bmatrix} K((x,c_1),(x',c_1)) & K((x,c_1),(x',c_2))&\cdots &K((x,c_1),(x',c_k))\\
        K((x,c_2),(x',c_1)) & K((x,c_2),(x',c_2)) & \cdots & K((x,c_2),(x',c_k))\\
        \vdots & \vdots & \ddots &\vdots\\
        K((x,c_k),(x',c_1)) & K((x,c_k),(x',c_2))&\cdots & K((x,c_k),(x',c_k))
    \end{bmatrix}\\
    & = \begin{bmatrix} \widetilde{K}_{11} & \widetilde{K}_{12}&\cdots &\widetilde{K}_{1k}\\
        \widetilde{K}_{21} & \widetilde{K}_{22} & \cdots & \widetilde{K}_{2k}\\
        \vdots & \vdots & \ddots &\vdots\\
        \widetilde{K}_{k1} & \widetilde{K}_{k2}&\cdots & \widetilde{K}_{kk}
    \end{bmatrix}\\
    & = \widetilde{K}(x,x').
\end{align*}

Then assume $\widetilde{Z}$ is a Gaussian $k$-variate random field on $\mathcal{Y}$. We prove that $Z\colon\Phi^{-1}(\widetilde{Z})$ is a Gaussian random field on $\mathcal{Y}\times\mathscr{C}$ with covariance function $K$. It suffices to check that $\cov(Z(x,c_i),Z(x',c_j))=K((x,c_i),(x',c_j))$ for any $x,y\in\mathcal{Y}$, $i,j=1,\cdots,k$. Then,
\begin{align*}
   \cov(Z(x,c_i),Z(x',c_j)) &=\cov(\widetilde{Z}_i(x)),\widetilde{Z}_i(x')))\\
   & = \widetilde{K}(x,x')_{ij}\\
   & = K((x,c_i),(x',c_j)).
\end{align*}
\end{proof}

\subsection{More covariance functions}\label{sec:covs}
We provide more semi-isotropic covariance functions below.
\begin{eqnarray}
 K((x,c_i),(x',c_j))&=&\frac{\sigma^2(a^2d_{ij}^2+1)}{[(a^2d_{ij}^2+1)^2+b^2\|x-x'\|^2]^{\frac{p+1}{2}}},\\
 K((x,c_i),(x',c_j))&=&\frac{\sigma^2(ad_{ij}+1)}{[(ad_{ij}+1)^2+b^2\|x-x'\|^2]^{\frac{p+1}{2}}},\\
 K((x,c_i),(x',c_j))&=&\sigma^2\exp\left\{-a^2d_{ij}^2-b^2\|x-x'\|^2-cd_{ij}^2\|x-x'\|^2\right\},\\
 K((x,c_i),(x',c_j))&=&\sigma^2\exp\left\{-ad_{ij}-b^2\|x-x'\|^2-cd_{ij}\|x-x'\|^2\right\}.
\end{eqnarray}

\subsection{Code}
All code for the models and experiments can be found in our GitHub repository: \url{https://github.com/andrewcharlesjones/multi-group-GP}. Within this repository, we provide an installable Python package for model fitting, covariance functions, prediction, and analysis.

\subsection{Data}
\subsubsection{GTEx}
\label{appendix:gtex_data}
The GTEx data can be downloaded from the GTEx portal: \url{https://gtexportal.org/home/datasets}. We use $52$ tissue types, listed below, although some experiments use a subset of these tissue types. The sample size for each tissue type is shown in square brackets.
Adipose Subcutaneous [644],
Adipose Visceral (Omentum) [539],
Adrenal Gland [254],
Artery Aorta [424],
Artery Coronary [238],
Artery Tibial [657],
Bladder [21],
Brain Amygdala [147],
Brain Anterior cingulate cortex (BA24) [172],
Brain Caudate (basal ganglia) [230],
Brain Cerebellar Hemisphere [208],
Brain Cerebellum [241],
Brain Cortex [255],
Brain Frontal Cortex (BA9) [200],
Brain Hippocampus [188],
Brain Hypothalamus [193],
Brain Nucleus accumbens (basal ganglia) [232],
Brain Putamen (basal ganglia) [194],
Brain Spinal cord (cervical c-1) [155],
Brain Substantia nigra [133],
Breast Mammary Tissue [456],
Cells Cultured fibroblasts [444],
Cells EBV-transformed lymphocytes [174],
Cervix Ectocervix [9],
Cervix Endocervix [10],
Colon Sigmoid [373],
Colon Transverse [406],
Esophagus Gastroesophageal Junction [375],
Esophagus Mucosa [551],
Esophagus Muscularis [515],
Fallopian Tube [9],
Heart Atrial Appendage [422],
Heart Left Ventricle [428],
Kidney Cortex [85],
Kidney Medulla [4],
Liver [224],
Lung [573],
Minor Salivary Gland [162],
Nerve Tibial [615],
Ovary [180],
Pancreas [37],
Pituitary [283],
Prostate [245],
Skin Not Sun Exposed (Suprapubic) [595],
Skin Sun Exposed (Lower leg) [665],
Small Intestine Terminal Ileum [187],
Spleen [233],
Stomach [359],
Testis [359],
Uterus [142],
Vagina [156],
Whole Blood [139].

\subsection{Supplementary figures}
\begin{suppfigure}[!h]
    \centering
    \includegraphics[width=\textwidth]{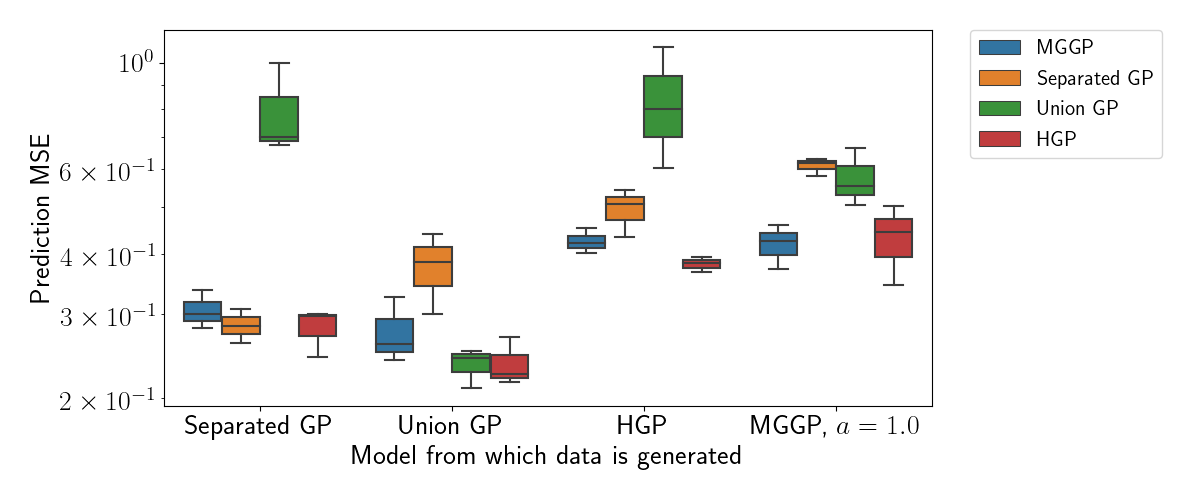}
    \caption{\textbf{Prediction experiment with synthetic data using a Matérn covariance function}. }
    \label{fig:prediction_simulation_experiment_matern}
\end{suppfigure}

\begin{suppfigure}[!h]
    \centering
    \includegraphics[width=0.9\textwidth]{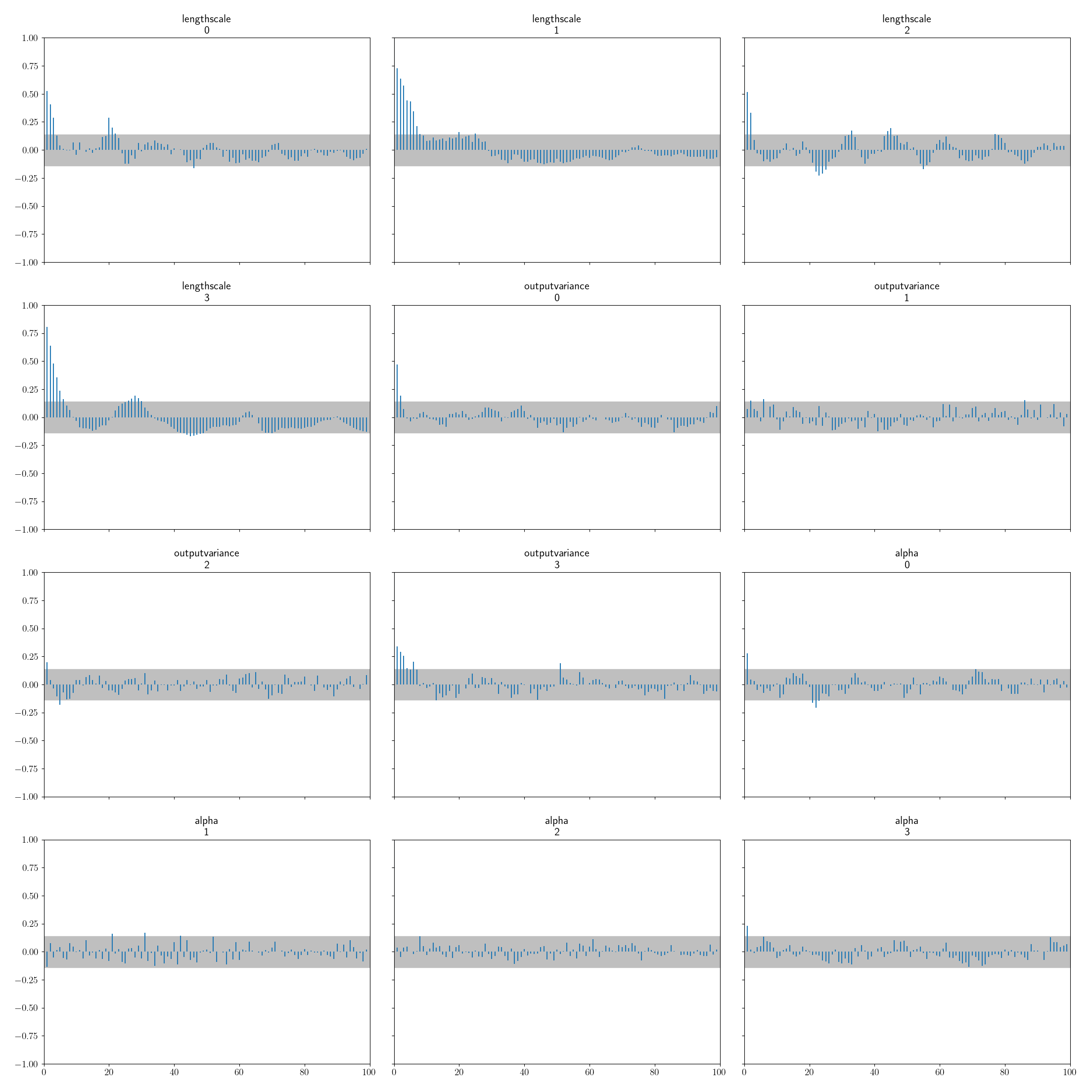}
    \caption{\textbf{Autocorrelation plots for MCMC samples from MGGP posterior}.}
    \label{fig:mggp_autocorrelation}
\end{suppfigure}

\begin{suppfigure}[!h]
    \centering
    \includegraphics[width=0.7\textwidth]{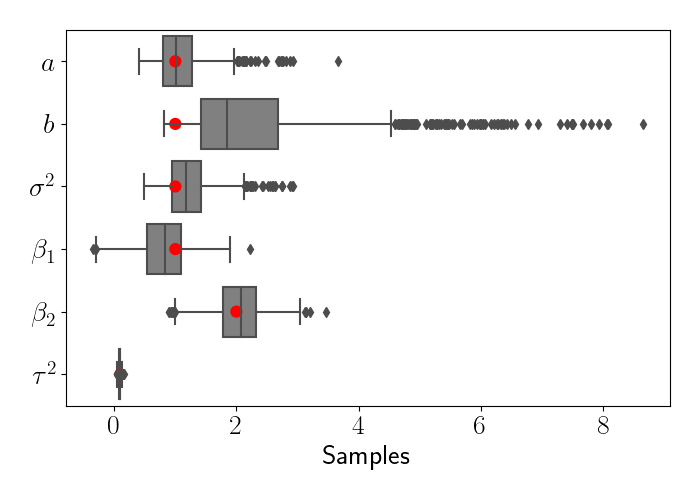}
    \caption{\textbf{Posterior samples of covariance function and model parameters}. Boxplots show the 95\% confidence intervals for each set of samples. Red points indicate the parameter values used to generate the data.}
    \label{fig:mggp_hyperparameter_samples}
\end{suppfigure}

\begin{suppfigure}[!h]
    \centering
    \includegraphics[width=0.7\textwidth]{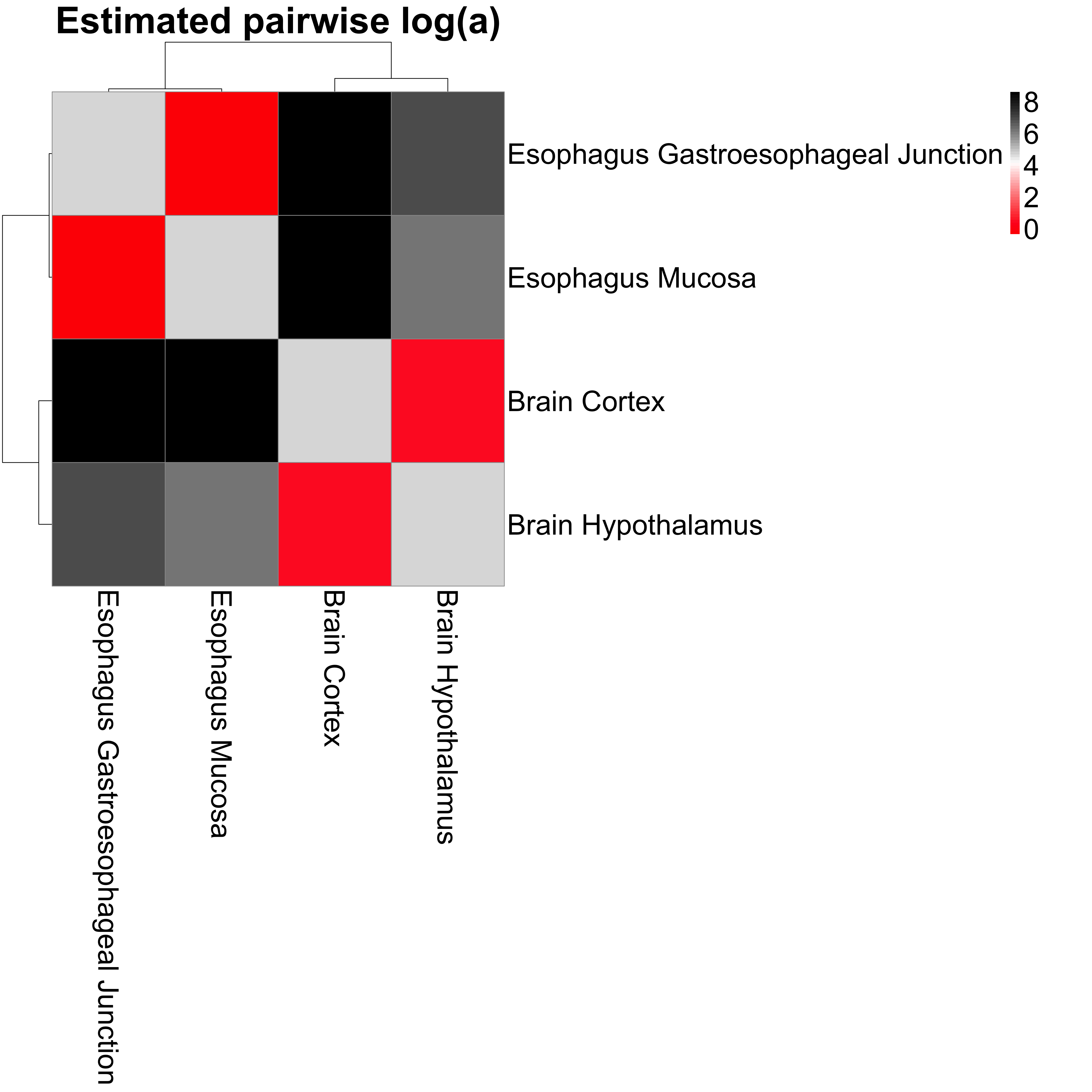}
    \caption{\textbf{Estimation of $a$ for each pair of GTEx tissue types}. Cell $ij$ in the heatmap represents $\log_{10}(a_{ij})$, where $a_{ij}$ is the maximum likelihood estimate of $a$ when fitting the MGGP using tissues $i$ and $j$. Lower values of $a$ (red) indicate higher similarity, while higher values of $a$ (black) indicate lower similarity. Here, we allow each group its own noise variance $\tau^2_j$.}
    \label{fig:gtex_pairwise_a_heatmap_small_group_specific_variances}
\end{suppfigure}

\begin{suppfigure}[!h]
    \centering
    \includegraphics[width=0.7\textwidth]{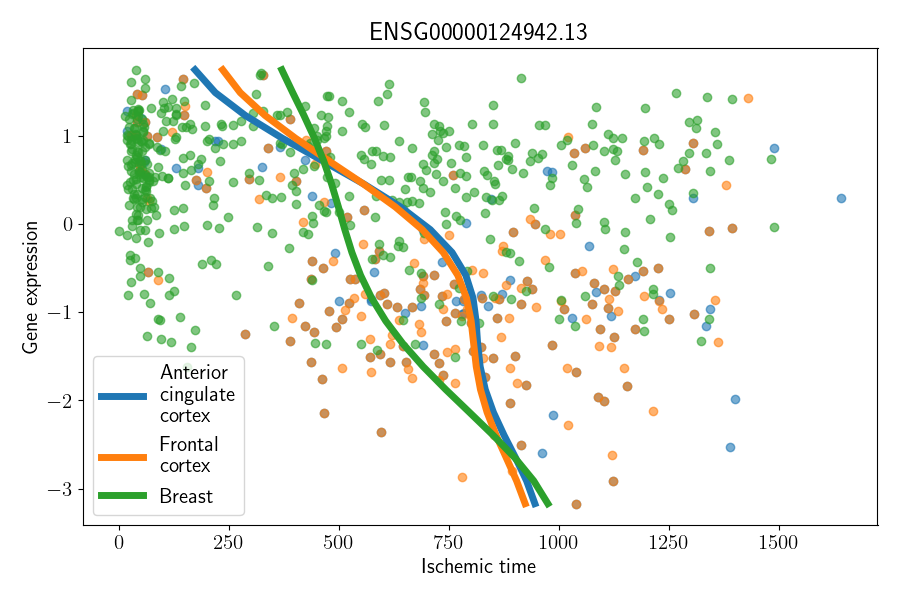}
    \caption{\textbf{Examining univariate relationships with the MGGP}. We reran the experiment shown in \autoref{fig:gtex_one_gene_relationship}, but here we used a noninformative distance matrix such that all three groups were equidistant.}
    \label{fig:gtex_one_gene_relationship_equidistant_ENSG00000124942}
\end{suppfigure}

\bibliographystyle{hapalike}
\bibliography{ref.bib}
\end{document}